\documentclass[12pt,preprint]{aastex}

\shorttitle{The NGC 300 2008 Optical Transient}
\shortauthors{Humphreys et al.}

\begin{document}

\title{The Photometric and Spectral Evolution of the 2008 Luminous Optical
Transient in NGC~300\altaffilmark{1}}

\author{
Roberta M. Humphreys\altaffilmark{2},
Howard E. Bond\altaffilmark{3}, 
Alceste Z. Bonanos\altaffilmark{5},
Kris Davidson\altaffilmark{2},
L. A. G. Berto Monard\altaffilmark{6},
Jos\'e L. Prieto\altaffilmark{7,9},
and
Frederick M. Walter\altaffilmark{8}
}

\altaffiltext{1}  
{Based in part on observations with the NASA/ESA {\it Hubble Space Telescope\/}
obtained at the Space Telescope Science Institute, and from the data archive at
STScI, which is operated by the Association of Universities for Research in
Astronomy, Inc., under NASA contract NAS5-26555; in part on archival data
obtained with the {\it Spitzer Space Telescope}, which is operated by the Jet
Propulsion Laboratory, California Institute of Technology, under a contract with
NASA; in part on observations obtained with the 6.5-m Magellan Clay Telescope
located at Las Campanas Observatory, Chile; in part on observations obtained
with the SMARTS Consortium 1.3- and 1.5-m telescopes located at Cerro Tololo
Interamerican Observatory, Chile, and in part on observations with VLT/UT1/ISAAC
under ESO DDT proposal: 283.D-5019.}

\altaffiltext{2}
{Minnesota Institute for Astrophysics, 116 Church St SE, University of
Minnesota, Minneapolis, MN 55455; roberta@umn.edu, kd@astro.umn.edu}

\altaffiltext{3}
{Space Telescope Science Institute, 3700 San Martin Drive, Baltimore, MD
21218; bond@stsci.edu}


\altaffiltext{5}
{Institute of Astronomy and Astrophysics, National Observatory of Athens, Athens
GR-15236 Greece}

\altaffiltext{6}
{Klein Karoo Observatory (formerly Bronberg), PO Box 281, Calitzdorp 6660, Western Cape, South Africa}

\altaffiltext{7}
{Carnegie Observatories, 813 Santa Barbara St., Pasadena, CA 91101}

\altaffiltext{8}
{Department of Physics \& Astronomy, Stony Brook University, Stony Brook, NY
11794-3800}

\altaffiltext{9}
{Hubble, Carnegie-Princeton Fellow}


\begin{abstract}

The 2008 optical transient in NGC 300 is one of a growing  class of
intermediate-luminosity transients that brighten several orders of
magnitude  from a previously optically obscured state. The origin of their
eruptions is not  understood. Our multi-wavelength photometry and spectroscopy
from maximum light to  more than a  year later provide a record of its
post-eruption behavior. We describe its changing spectral-energy distribution, 
the evolution of its absorption- and emission line-spectrum, the development of
a bipolar outflow, and the rapid transition from a dense wind to an  optically
thin ionized wind.  In addition to strong, narrow hydrogen lines, the F-type
absorption-line spectrum of  the transient is characterized by strong
\ion{Ca}{2} and [\ion{Ca}{2}] emission. The very broad  wings of the \ion{Ca}{2}
triplet and the asymmetric [\ion{Ca}{2}] emission lines are  due to  strong
Thomson  scattering in the expanding ejecta.  Post-maximum, the hydrogen and
\ion{Ca}{2} lines developed double-peaked emission profiles that we attribute to
a bipolar outflow. Between  approximately 60 and 100 days after maximum, the
F-type absorption spectrum, formed  in its dense wind, weakened and the wind
became transparent to ionizing radiation. We discuss the probable evolutionary
state of the transient and similar objects such as SN~2008S, and conclude that
they were most likely post-red supergiants or post-AGB stars on a blue loop to
warmer temperatures when the eruption occurred. These objects are not LBVs.  

\end{abstract}


\section{Introduction: The Intermediate-Luminosity Transients}

The numerous surveys for supernovae and transient variables have identified several non- terminal eruptive objects with maximum luminosities between novae and supernovae. Some have received 
a supernova designation (SN) and consequently have been called ``supernova 
impostors'' \citep{vandyk}. 
Many of these objects appear to be the ``giant eruption'' of a massive hot star perhaps 
similar to $\eta$ Car (Humphreys \& Davidson 1994, Humphreys et al. 1999). A few, like the 
2008 optical transient in NGC 300  \citep{Bond09,Berger09}, discussed here, are apparently 
the unexplained outburst or eruption of a  dust enshrouded progenitor of  intermediate luminosity and mass, $\sim$ 10 -- 17M$_{\odot}$ \citep{Bond09,Gog09}.  
 Prieto et al. (2009) have reported the presence of broad  emission features in the 
 mid-infrared spectrum of the transient similar  to features seen in carbon-rich proto-planetary 
  nebulae supporting their suggestion that the progenitor is actually a somewhat lower mass 
  post-AGB star.

Related objects include SN~2008S \citep{Prieto08A,smith08,Thompson08} in
NGC~6946, SN~2010da \citep{Monard,Khan,Chornock,Bond10} also in NGC~300,
SN~2010dn in NGC~3184 (Smith et al.\ 2011), and the so-called ``luminous red
novae'' PTF~10fqs (Kasliwal et al.\ 2011) in M99  and PTF~10acbp in UGC 11973
(Kasliwal et al.\ 2010). The red transient in M85 \citep{Rau07,Kulk07} and the
M31 red variable, M31~RV (see Bond \& Siegel 2006; Bond 2011; and references
therein), may have been similar outbursts,  although they  occurred in older
stellar populations and presumably from lower-mass stars. The recently announced
PSN~J17592296+0617267 in NGC~6509 (Kelly et al.\ 2011) appears to be very
similar to SN~2008S and the 2008 NGC~300 transient.  These  transients  thus
occur in intermediate-type and cool evolved stars and possibly over a relatively
wide mass range. 

Given the very small sample, it is unclear if the causes of their instabilities
and  eruptions are physically related. A second group of  transients  may be
more closely related to binary-star interactions. For example, the
well-studied,  peculiar eruptive  star, V838 Monocerotis, famous for its
spectacular light echo \citep{Bond03},  was  very different from a SN impostor
or LBV-type eruption.  Although V838~Mon  later became enshrouded in dust, its
light curve was also not like the other transients.   The cause of its eruption
is not known, but because of its unique properties, it has been speculated that
it may have been the result of a stellar merger \citep{TS06}. Two other Galactic
examples are the nova-like V4332~Sgr \citep{Martini} and V1309~Sco \citep{Mason}, which were
both relatively red and cool during their outbursts. 

Even though these intermediate-luminosity transients  are rare, with the
occurrence of a second, similar object  in NGC~300 (SN~2010da) and the recent discovery that
the outburst of V1309~Sco was caused by the merger of a contact binary
\citep{Tylenda}, there is increased   interest in these objects, the origins of
their outbursts, and the nature of the progenitors. Bond (2011) has suggested  calling all 
of these
objects ``intermediate luminosity red transients'' (ILRTs) to distinguish them from true novae and LBVs.

This paper presents a detailed photometric and spectroscopic investigation of
the 2008 NGC~300 optical transient (OT), ``NGC~300 OT2008-1''.   
 Hereafter in this paper we refer to it simply as ``the OT'' or ``the transient''. 
 In an earlier {\it Letter\/}
\citep{Bond09}, we presented a brief overview of its light curve and spectral
characteristics. Berger et al.\ (2009)  also published a 
discussion of   their post-maximum spectral series in which they  proposed a
complex model for the its circumstellar ejecta and eruption.   To provide more
information on its post-eruption behavior, we describe here its spectral energy
distribution,  dust formation, and the evolution of the optically thick wind
and  outflow from maximum to late times.  Our multi-wavelength observations and
the light curve are presented  in the next section. The color-magnitude diagram
and stellar-population characteristics in the OT's immediate environment are
described in \S {3}. Its changing spectral energy distribution and dust
formation, the energetics of the eruption, and  the luminosity and mass of the
progenitor are  discussed in \S {4}.   In \S {5} and {6} we describe the
behavior of the emission and absorption lines as the eruption declined, the
development of the bipolar outflow, and the  transition to an optically thin
wind. In the final section we discuss some possible  explanations for the
outburst and comment on the transient's evolutionary state.

\section{Multi-wavelength Observations and Data Reduction}

Our observations include ground-based CCD imaging and
photometry, near-infrared photometry, low- to moderate-resolution optical and
near-infrared spectroscopy, {\it Hubble Space Telescope\/} ({\it HST\/})
imaging,   and {\it Spitzer\/} mid-infrared photometry (Prieto 2008). We
conducted  a program of photometric monitoring of the  transient from 2008 May
15 until 2009 June 22, and spectroscopy from 2008 May 15 to 2009 January  16
with the 1.3- and 1.5-m telescopes at Cerro Tololo Interamerican Observatory
(CTIO), operated by the SMARTS consortium\footnote{SMARTS is the Small and
Medium Aperture Research Telescope System; see
http://www.astro.yale.edu/smarts/}. These spectra were supplemented by
echellette spectra from the Magellan telescope and near-infrared spectra from
the VLT.  

\subsection{Optical and Near-Infrared Imaging and Photometry}

The outburst was discovered by Monard (2008) at broad-band
magnitude 14.3, using his 0.3-m telescope equipped with a CCD camera, on a frame
obtained on 2008 May 14 (UT)\null. An earlier frame, taken on 2008 April 24, and
examined after discovery of the eruption, showed the transient at
$\sim$16.3~mag. It was fainter than 18~mag on 2008 February~8, and on all
previous monitoring observations obtained by Monard.

Our program of systematic photometric monitoring began on 2008 May 15. These
data were obtained by Chilean service observers with the 1.3-m SMARTS telescope
at CTIO with the ANDICAM optical/near-IR direct camera (DePoy et al.\ 2003),
which simultaneously obtains optical frames with a CCD detector and near-IR
frames with an IR detector. We used the {\it BVRI\/} filters of the
Johnson-Kron-Cousins system in the optical, and standard {\it JHK\/} filters for
the near-IR\null.

The {\it BVRI\/} frames were processed through the SMARTS reduction pipeline at
Yale University, which performs bias subtraction and flat-fielding. We then used
standard aperture-photometry routines in  IRAF\footnote{IRAF is written and
supported by the IRAF programming group at the National  Optical Astronomy
Observatories (NOAO) in Tucson, Arizona. NOAO is operated by the  Association of
Universities for Research in Astronomy (AURA), Inc. under cooperative agreement
with the National Science Foundation} to determine magnitudes differentially
with respect to a nearby comparison star. The comparison star is located at
J2000 coordinates (USNO-NOMAD catalog) RA = 00:54:39.545, Dec = $-37$:37:14.79. 
The magnitudes of the comparison star were calibrated on seven photometric
nights by reference to standard stars  published by Landolt (1992), and are
$V=16.07$, $B-V=0.75$, $V-R=0.42$, and $V-I=0.82$ with errors of 
$\pm$0.02--0.03~mag.   The {\it JHK\/} magnitudes are also differential with
respect to the same comparison star, calibrated using its 2MASS photometry. The
J2000 position of the OT, based on astrometry of ANDICAM frames calibrated
against the USNO-NOMAD catalog, is RA = 00:54:34.51, Dec = $-37$:38:31.4, with
errors in each coordinate of $\pm$$0\farcs2$.

The journal of observations and the multi-wavelength photometry is in Table~A1
in the Appendix in the electronic edition.  Figure 1 shows the {\it BVRIJHK\/}
light curve. The rise to maximum was poorly covered, but it was much more rapid
than the subsequent slower decline. The brightest $V$ magnitude we measured was
14.69, on the first night of SMARTS observations (2008 May 15), corresponding to
an absolute visual magnitude $M_{V,0}=-12.2$ to $-13.1$ for the adopted distance
(see \S {3}) and reddening in the range $E(B-V)=0.1$ to $0.4$ (see \S {4}).
Following maximum light, the transient  declined smoothly in brightness at all
wavelengths, but with a noticeable change in slope in 2008 September. The $R$
magnitudes did not fade as rapidly as in the other bandpasses, due to strong
H$\alpha$ emission.

In Figure 2 we plot the changes in the color indices $B-V$ and $V-K$ with time.
The $V-K$ color became steadily redder as the transient  evolved, but with a
change to a shallower slope in 2008 September due to the change in slope of the
$V$ magnitude. $B-V$ also became redder (but note the expanded scale for this
color in Fig.~2) until 2008 August, but then trended back toward bluer colors.
These changes in color are discussed further in \S {4}.

\subsection{Spectroscopy}

We regularly obtained low- and moderate-resolution spectra of the transient
throughout its outburst from 2008 May 15 to 2009 January 16, using the long-slit
Ritchey-Chretien (RC) spectrograph on the SMARTS 1.5-m. We used a $1\farcs5$
slit and a variety of gratings, yielding spectral resolutions from 3.1 to
17.2~\AA, and covering the spectral range 3150 to 9350~{\AA} (see Table~1). Each
 observation consisted of three integrations, which were median
filtered to remove cosmic rays  and then summed. Each set of images was
wavelength calibrated with a Ne-Ar or Th-Ar arc lamp.  The images were
bias-subtracted, trimmed, and flattened using dome flats obtained each
night.  The spectra were extracted by fitting a Gaussian in the spatial
dimension at each column in the CCD\null. The net counts at each pixel are the
integrated  counts in the Gaussian, less the interpolated background fit on
either side of the spectrum. The  spectrophotometric standard, Feige~110, was
observed each night to  convert the counts to flux.  Due to seeing-related slit
losses and possible transparency changes   during the night we do not obtain
absolute fluxes, but rather merely recover the shape of the continuum. We
correct to true fluxes by scaling them to match the observed $BVRI$ magnitudes,
interpolating as necessary. The observation dates and grating used are
summarized in Table~A2 in the Appendix in the electronic edition.  

We also obtained moderate-resolution spectra with the Magellan 6.5-m telescope
and MagE echellette spectrograph (Marshall et al.\ 2008) on  2008 July 6,
August~30, and September 1.  These spectra cover the wavelength region from
3100~{\AA} to $1\,\mu$m. The integration times were $2 \times 800$~s on July 6
and 900~s each on August 30 and September~1.  The $0\farcs7$ slit yielded a
resolving power $R\simeq6000$, corresponding to a resolution of 1.5~{\AA} at the
\ion{Ca}{2} triplet and 1.1~{\AA} at H$\alpha$. The MagE spectra were reduced,
extracted and wavelength calibrated with an IDL pipeline written by G.~Becker
(version 1.01), which implements optimal extraction techniques (e.g., Kelson
2003). A fourth MagE spectrum was obtained on 2009 June 5, more than a year 
after the eruption.  With a combined 3000~s integration, there is no continuum
and the only emission lines detected are H$\alpha$ and the \ion{Ca}{2}
triplet.  

We also obtained Director's Discretionary Time at the European Southern
Observatory for near-infrared spectroscopy   with the  Infrared Spectrometer and
Array Camera (ISAAC) on the VLT/UT3 telescope on 2009 July 19--23, more than a
year after maximum.  These low-resolution spectra  were observed at central
wavelengths 1.65~$\mu$m, 2.2~$\mu$m, and 3.55~$\mu$m, and a medium-resolution
spectrum was obtained at 2.2~$\mu$m. Due the faintness of the object (see
Table~A1), the spectra have low S/N\null. Br$\gamma$ and Pa$\alpha$ are present
in emission. 

\section{The Stellar Environment}

As we described in Bond et al.\ (2009), the {\it HST\/} archive contains two
sets of deep frames made before the outburst  that fortuitously cover the OT
site, obtained with the Wide Field Channel (WFC) of the Advanced Camera for
Surveys (ACS) and the F435W, F475W, F555W, F606W, and F814W filters. These
observations were made on 2002 December~25 (GO-9492, PI: F.~Bresolin) and on
2006 November 8--10 (GO-10915, PI: J.~Dalcanton). The frames were obtained in
three different pointings in a spiral arm of NCG~300 lying to the northwest of
the nucleus. Field~5 in Bresolin et al.\ (2005) includes the site of the
transient,  one of the Dalcanton frames largely overlaps it, and a second one is
adjacent to it on the SE side, as shown in Gogarten et al.\ (2010, their
Fig.~1).  We also obtained two observations during the  outburst in our {\it
HST\/} Director's Discretionary (DD) program GO-11553 (PI: H.~Bond). The DD
observations were made on 2008 June~9 and September~1, using the Wide Field
Planetary Camera~2 (WFPC2) with F450W and F814W filters. We used these data to
locate the OT site precisely in the pre-eruption frames and search for a
progenitor object. Remarkably, no progenitor star was detected in any of the ACS
frames, to a magnitude limit of 28.5 in the F606W image.

These deep pre-outburst frames can be used to study the surrounding stellar
population, and thus make inferences about the possible nature of the progenitor
star. We made brief comments about the stellar environment in Bond et al.\
(2009), including the fact that there are several blue main-sequence stars
within $2\farcs5$ (23~pc) of the OT site with initial  masses as high as
$\sim$$14\,M_\odot$\null. An extensive study of the stellar populations and
star-formation history around the OT was made by Gogarten et al.\ (2009), based
on their independent analysis of the same  {\it HST\/} material. Due to the
presence of stars in the immediate vicinity of the OT with ages of only
$\sim$8--13~Myr, they concluded that the progenitor was likely to have had an
initial mass in the approximate range 12--$17\,M_\odot$.

Our independent photometric analysis of the archival {\it HST}/ACS material
was done  using techniques described in detail by Anderson et al.\ (2008).
Because our findings largely verify those reported by Gogarten et al.\ (2009),
we give only a brief summary here. We prepared stacked images by combining all
of the available ACS frames. The combined images cover slightly more than the
equivalent of two ACS/WFC fields, each of which is $202''\times202''$ or about
$1.8\times1.8$~kpc at the distance of NGC~300. We then carried out stellar
photometry on these combined images, calibrated  on the Vega-mag system, as
described by Bedin et al.\ (2005), and adopting the ACS/WFC encircled-energy
function and photometric zero-points given by Sirianni et al.\ (2005).

Figure 3 shows the resulting color-magnitude diagrams (CMDs) for all stars
detected in the stacked ACS frames (the black points in the diagrams). The
left-hand frame shows $I$ vs.\ $V-I$ ($m_{\rm F814W}$ vs.\ $m_{\rm F606W}-m_{\rm
F814W}$), and the right-hand frame shows $I$ vs.\ $B-I$ ($m_{\rm F814W}$ vs.\
$m_{\rm F475W}-m_{\rm F814W}$)\null.   The CMDs reveal a combination of a young
population (the nearly vertical plumes of blue main-sequence stars, blue loop
stars, and red supergiants), an intermediate-age population (the AGB stars lying
above the red-giant tip), and an underlying older population of numerous red
giants.  

The tip of the red-giant branch (TRGB) is well-defined in Figure 3, and allows
us to determine the distance to NGC~300 using this well-known method (e.g.,
Madore \& Freedman 1995). We first transformed our data onto the
Johnson-Kron-Cousins $VI$ system, using relations given by Sirianni et al.\
(2005), and then determined the luminosity function (LF) of the RGB stars. We
then applied a Sobel filter to the LF, using techniques described by Sakai,
Madore, \& Freedman (1996), to determine the $I$ magnitude of the TRGB\null. We
find $I_{\rm TRGB}=22.60 \pm 0.10$, with the quoted error being the HWHM of the
Sobel peak. Adopting $E(B-V)= 0.037$ from Gogarten et al.\ (2009) and $A_I=2.0
E(B-V)$, we obtain $I_{0,\rm TRGB}=22.53 \pm 0.10$, including a $\pm$0.02~mag
uncertainty associated with the reddening estimate. Adopting an absolute
magnitude $M_{I,\rm TRGB}=-4.04\pm0.12$ (e.g., Rizzi et al.\ 2006), we derive a 
distance modulus of $(m-M)_0 = 26.57\pm 0.14$, corresponding to 
$2.03\pm0.13$~Mpc. Our result is in good ($1\sigma$) agreement with the distance
modulus adopted in an extensive discussion of the NGC~300 literature by Gogarten
et al.\ (2010), $(m-M)_0 = 26.43 \pm 0.09$ (1.93~Mpc).

To better characterize the stellar population in the vicinity of the OT, we
consider only the stars located within a $500\times500$~pc square centered on
its position. The CMD for these stars is shown in Figure~4 (left panel), with
stars lying in different evolutionary states indicated by the color-coding, as
described in the caption. In the right panel, we show the spatial distribution
of the various types of stars by zooming in on a $200\times200$~pc square with
the same color-coding.   

This diagram reveals several features. First, the entire field is underlain by a
substrate of old red giants (dark-red points). There is also a fairly uniform
scattering of blue main-sequence and blue-loop stars (slate-blue and bright-blue
points) across the field, but with a slight gradient toward higher stellar
densities to the west (right-hand) side of the field. There is little evidence
among these blue stars for clustering or young associations, although there {\it
are\/} several obvious clusters and associations not far outside the field in
the west and southwest direction. Several of the young main-sequence stars lie
within $\sim$$2''$--$3''$ ($\sim$18--27~pc) of the transient's  site, as already
noted by Bond et al.\ (2009) and Gogarten et al.\ (2009). The rarer red
supergiants (orange points) and AGB stars (bright-red points) are distributed
approximately uniformly across the field. This figure thus reveals a mixed
stellar population in the transient's immediate vicinity. 

\section{The Spectral Energy Distribution, Energetics of the Eruption, and
Luminosity and Mass  of the Progenitor}

Figures 5 and 6 show the transient's spectral energy distribution (SED) at
maximum light and its subsequent evolution from maximum to 2009 June 22, more
than a year later.   

The observed $BVRI$ and $JHK$ magnitudes from day 2 in the eruption are shown in
Figure~5.  The same magnitudes are plotted corrected for the mean interstellar
extinction, $A_{V}\approx 0.3$~mag, for NGC~300 \citep{Gieren05} and from the
field stars  near  the transient \citep{Gog09}.   We also show the photometry
corrected for $A_{V} = 1.2$~mag assuming  an intrinsic color ($B-V_{0}\simeq
0.3$ to 0.5~mag) appropriate to its F-type  absorption spectrum at maximum
\citep{Bond09}.  $E(B-V)$ is then $\sim$0.4~mag from its observed $B-V=0.8$ at
maximum. The  7500~K blackbody, fit through the latter points,  is consistent
with  the  F-type absorption spectrum  observed at maximum and for several
months afterwards. The absorption spectrum,  discussed in \S {5} and {6}, very
likely originates in a dense wind, but could also   correspond to the
progenitor's actual photosphere.  In either case, 7500~K is the apparent
temperature of the region from which the visual  light is escaping. An energy
distribution of comparable temperature cannot be fit to the photometry corrected
for the lower extinction value. A much lower temperature of  4000--4500~K,
inconsistent with the spectrum, would be necessary to fit the $B-V$ color
corrected for the lower extinction,  but even then it is not a good fit to $B$
and does not match the $R$ and $I$ magnitudes.  We therefore adopt $A_{V}=
1.2$~mag for this discussion. This higher  $A_{V}$ may simply be due to higher
interstellar reddening for the  OT than for the surrounding field stars, but as
we show here, it is more likely due to circumstellar extinction from dust.

In Figure~5, a 7500 K blackbody  fit to the $B$, $V$, and $R$  magnitudes 
leaves a small excess at $I$ and $J$  above the blackbody which, at these
wavelengths, is due to free-free emission in the star's wind. The contribution
from a 7500~K free-free  curve is also shown.  The adopted temperature for the
free-free emission is somewhat  arbitrary, but it does not make much difference;
a $T_{\rm ff}$ in the  5000--10000~K range  would look very similar. The
combined blackbody plus free-free fit leaves a small  excess at $H$ and a
large excess at $K$, presumably due to dust. If we
assume that the correction  due to interstellar extinction is 0.3 mag and  the
0.9 mag difference is circumstellar,   then about half of the total flux is
absorbed by dust.   We can therefore model the dust contribution assuming that
the integrated flux from dust is 50\% of the blackbody plus free-free flux, and
the combined flux plus dust must match  the $K$-band magnitude. These conditions
are fulfilled by the 715~K dust contribution  shown in Figure~5. The resulting
SED matches the $BVRI$ and $JK$ fluxes very well. The $H$-band magnitude was not
used in this fit, but is also matched quite well.   This result suggests that
some dust either survived the eruption, or was formed during the
outburst. 

With an $A_{V}$ of 1.2~mag, the luminosity of the OT at maximum was  $1.5 \times
10^{7} \, L_{\odot}$ with a photospheric radius of $\sim$11~AU for an apparent temperature  
 of 7500~K\null. The 715~K dust is at  about 1500~AU (9  light days) from
the star.  The  $K$-band photometry shows considerable variability between day
2  and day 33 (Figure~2), as much as 0.2 to 0.3 mag in one day (see Table~A1).
Although it  is tempting to  attribute this to variable dust  formation
immediately after the eruption, variations  of this  size in less than a week
would be hard to explain. Furthermore,   with expansion velocities of 500--600
km s$^{-1}$ at maximum \citep{Berger09}, the ejecta  from the eruption would
require more than a year to reach this distance.  {\it Therefore, the dust must
be a remnant from the star's previously obscured state}.  In their discussion of
the carbon-rich mid-IR emission feature and the SED at day 93 (see below),
Prieto et al.\ (2009) also concluded that the corresponding dust survived the
eruption. 

Figure 6 shows the observed SED at several different times after maximum, along
with  the pre-outburst  infrared SED \citep{Prieto08} and the mid-IR emission
feature observed on 2008 August 14  \citep{Prieto09}. As the OT faded, it also
became redder, but its characteristic F-type absorption-line  spectrum was still
weakly present  in mid-August 2008 (day 90); see \S {5} and \S {6}.    We
therefore attribute the redder color to increased dust formation and  
circumstellar reddening.  Following the discussion above, we estimate the
combined interstellar and circumstellar extinction from the observed $B-V$ color
and F-type spectrum to determine the extinction-corrected SEDs also shown in
Figure~6. The adopted values of $A_{V}$ are included in the figure caption.  By
2008 September (day 117) the absorption spectrum was clearly gone and  the $B-V$
color was bluer, although $V-K$ was redder.  We adopted the last value of  
$A_{V}$ for the later dates.  After that, the SED rapidly becomes  much steeper
in the visual.  The change in the shape of the SED, the disappearance of the
absorption spectrum,  and the bluer $B-V$ correspond to the change in the slope
of the visual light curve in Figure 1   beginning in 2008 September.  These
changes are very likely due to the  transition in the dominant radiation field
described  in \S {5} and \S {6}.  By the time of our  last visual measurement,
day 405, the transient was still more luminous than the pre-outburst {\it
Spitzer\/} data \citep{Prieto08} shown in Figure~6. 

Its pre-eruption infrared SED indicates that the optically obscured progenitor
had an intrinsic luminosity of $6.3 \times 10^{4} L_{\odot}$, at our adopted
distance (2.02~Mpc) for  NGC~300, corresponding to  $M_{\rm bol} \simeq-7.3$,
placing it just above the AGB limit at $M_{\rm bol} \simeq-7.0$. At maximum
light, with its F-type absorption spectrum probably  due to the formation of an
optically thick  wind or envelope, its bolometric correction  will be near 
zero, so that $M_{V}\simeq M_{\rm bol}$ at maximum. With an $M_{V}$ of
$-13.1$ ($A_{V}\approx 1.2$) at maximum\footnote{With $A_{V} = 0.3$, $M_{V}$ at
maximum was $-12.2$.}, the OT thus increased its total luminosity by  about 
6~mag or a factor of 250, during its eruption.

The energy released in the  eruption, however, was relatively modest
compared to other non-terminal eruptions. Integrating over the visual light
curve, corrected for extinction ($A_{V}\simeq0.3$ to 1.2 mag), beginning with
the initial detection on 2008 Apr 24, and assuming that the bolometric
correction is zero, we find that the total emitted energy was
1--$2\times10^{47}$~ergs. Almost all of this energy was released in the first
two months. This is considerably less than the luminous energy,
$\sim$$10^{50}$~ergs, emitted in $\eta$~Car's famous great eruption, 
but it may be comparable to its second, lesser eruption
in the 1890's, which released $6 \times 10^{47}$~ergs during its 7-year
outburst. If the ratio of luminous to kinetic energy is near unity, as in
$\eta$~Car, then with a wind speed of $\sim$75--100 km s$^{-1}$ (see \S 6), the
total mass lost by NGC~300 OT2008-1 was 1--$2\, M_{\odot}$. However, this is
very likely an upper limit.  For example, in the  classical LBV ``eruptions''
the luminous energy is 10--100 times greater than the kinetic energy. A
realistic estimate of the mass shed in the eruption is
therefore somewhere between $\sim$0.1 and $1\,M_{\odot}$.

Comparison with evolutionary tracks for solar metallicity, with and without
rotation \citep{MM00}, suggests an initial mass of 10--$15\,M_{\odot}$ for the
progenitor star based on its pre-outburst luminosity, assuming
that it is an evolved red supergiant (RSG) or post-RSG star \citep{Bond09}. Note
that the progenitor is just the above the AGB limit. Using color-magnitude
diagrams to age-date the surrounding population, Gogarten et al.\ (2009) 
concluded  that the progenitor's initial mass was between 12 and
$17\,M_{\odot}$. Our inspection of the spatial distribution of the different
stellar populations in the  immediate neighborhood of the progenitor in \S {3}
is inconclusive. It could have originated from a fairly massive star or from an
intermediate-mass star of $\sim$5--$9\,M_{\odot}$  on the AGB\null.  The
detection of mid-IR emission features attributed to hydrocarbons (Prieto et al.\
2009) may suggest that the progenitor was a post-AGB star, similar to
proto-planetaries, and may have evolved from lower masses.

\section{Evolution of the  Spectrum}

The distinguishing characteristics of the transient's optical spectrum are its strong, 
narrow H$\alpha$ emission line,  strong  \ion{Ca}{2} triplet emission, and  the
rare [\ion{Ca}{2}] emission lines near  $\lambda$7300~{\AA}\null.   Our first
spectrum, obtained at maximum light, is shown in Figure~2 in our earlier {\it
Letter\/} (Bond et al.\ 2009).  Although the resolution of this first spectrum
is relatively low (17~{\AA}), it also shows several  absorption lines that
resemble the spectrum of a luminous F-type supergiant with strong H and K lines,
a weak G-band, \ion{Na}{1} D line blend, and absorption lines,  especially
around 4175~{\AA} and 4144~{\AA} that are  blends of lines that are strong in
F-type supergiants.  The \ion{O}{1} blend at 7774~{\AA}, a strong luminosity
indicator for A- and F-type supergiants \citep{Osmer}, is also present in
absorption.

An apparent F-type supergiant spectrum is expected in an eruption
that produces an optically thick wind or envelope.  Suppose that the star's
basic radius, in the absence of a wind, corresponds to an effective temperature
above $\sim$8000~K,  with a fixed luminosity.  If a sufficiently dense outflow
arises, the wind becomes opaque, so that the effective photosphere is then
located well outside the star. Under  these conditions, of course an increase in
the wind density tends to enlarge the  photospheric radius, thereby reducing the
apparent temperature.  However, the relevant  opacity declines precipitously
below 7500~K; therefore a photospheric temperature below  7000~K requires a very
high wind density.  This behavior was described by Davidson (1987),  who
emphasized that a broad range of mass-loss rates (even in the logarithmic
sense)  all give apparent temperatures between 6500 and 7500~K\null.  Since the
effective gravity is near zero, the resulting spectrum resembles an
intermediate-temperature supergiant.   This is why  LBVs/S~Doradus variables at
maximum have apparent A- to F-type supergiant absorption spectra
and energy distributions \citep{HD94}. 

Thus, the F-type absorption spectrum is most likely produced in the transient's
dense wind or it could  represent the progenitor's photosphere.  The absorption
lines do weaken with time and are eventually gone by  later times. These
apparent changes in  what may be a cool dense wind are discussed later. 

\subsection{The Hydrogen and \ion{Ca}{2} Emission-line Spectrum}

The spectrum is dominated by a very strong H$\alpha$ emission line.  Its 
equivalent width  increased by more than a factor of 20 from maximum light
through 2008  December, due   primarily to the  decreasing level of the
continuum as the object faded. The integrated line flux, however, shows that the
strength of H$\alpha$ did indeed increase, reaching a maximum 50--60 
days after maximum light (2008~July),  and  declined thereafter. At maximum
light, 2008 May 15, the H$\alpha$ profile had a FWHM of 1050~km s$^{-1}$, but
became narrower with time, and a year later, in 2009 June, the FWHM was only 300
km s$^{-1}$. The later H$\alpha$ profiles observed with our highest-resolution
RC grating and the  echellette spectra also show relatively symmetric, prominent
Thomson-scattering wings. These are produced in the  wind, with wavelengths 
 corresponding to velocity shifts  of the order of $\pm$900--1000 km
s$^{-1}$ (Figure~7),  due to scattering off the
electrons, not to bulk Doppler motions.  The scattering wings  may always have
been present in the early spectra, but only became apparent  as the continuum
level decreased. 

The \ion{Ca}{2} triplet and the  unusual [\ion{Ca}{2}]
emission lines are present in all of our spectra. These lines are indicative of
an  extensive circumstellar medium  and provide information on conditions in
the   ejecta, which, as we describe here, changed with time as the eruption
declined. The integrated fluxes for  the triplet lines noticeably increase in
strength, reaching  a maximum in 2008 July, like H$\alpha$, while the
integrated  fluxes for the [\ion{Ca}{2}] lines  continuously  decrease in
strength from May to December. The \ion{Ca}{2} triplet emission lines are formed
in the star's ejecta   by radiative de-excitation from the strong  \ion{Ca}{2} H
and K absorption upper levels, and are observed in the spectra of the
winds or envelopes of the warm  hypergiants,  but are not commonly observed in the 
spectra of the dense  winds of LBVs at maximum, their optically thick wind
state.   \ion{Ca}{2} has a low ionization potential and is suppressed in the
presence of UV radiation. The transition  that produces the triplet emission
leaves the ions in the upper level for the [\ion{Ca}{2}] forbidden lines. These
ions are  normally collisionally de-excited back to the ground state that
produces the H and  K lines, unless the density is sufficiently low.  For these
reasons the [\ion{Ca}{2}] emission lines are rarely seen in stellar spectra. 

We can  estimate the number  of photons that are radiatively de-excited  from
the ratio of the combined equivalent widths (or fluxes) of the [\ion{Ca}{2}]
lines and  the \ion{Ca}{2} triplet, multiplied by the   expected ratio of the
corresponding continuum fluxes  at 7300~{\AA} and 8600~{\AA},  respectively. 
For this calculation, we assumed a 7000~K blackbody.  At maximum  light in 2008
May,  we find that about 1/4 of the photons produced the forbidden lines by
radiative  de-excitation to the ground state. This is  similar to the results
for  the ejecta and winds  of the much more luminous stars IRC~+10420 (Jones et
al.\ 1993;  Humphreys et al.\ 2002)   and Var~A in M33 \citep{RMH06}. The
[\ion{Ca}{2}] lines  are thus produced in a region  of much lower density than
normally found in the  atmospheres and winds of luminous supergiants, and 
therefore  originate  outside  the  dense wind that formed during the eruption.
But that changed with time.  In later spectra, the ratio declined to one-tenth
in  2008 August through September  and to approximately one-twentieth by 2008
November.  This rapid decline in the photon ratio could be due to an increase in
the density in the ejecta where the [\ion{Ca}{2}] lines are formed, or to an
increase in the ionizing flux, as discussed below  and in \S {6}. 

Prieto et al.\ (2009) and Smith (2009) have suggested that the [\ion{Ca}{2}]
emission lines are  instead due to dust  destruction, which presumably populates
the necessary upper levels. However, the spectrum  of the transient has very
strong \ion{Ca}{2} H and K absorption lines which will populate the  upper
level, and so far, no one has done the necessary analysis to demonstrate that
dust  destruction will do likewise. 

\subsubsection{The Double--Peaked Emission-Line Profiles}

In the post-maximum spectra both the \ion{Ca}{2} triplet and H$\alpha$ show
prominent double or split emission lines with asymmetric profiles (see Figures~7
and~8).   Double-peaked  emission is normally  attributed to either a bipolar
outflow or a rotating disk. In this case, with an ongoing eruption, they are
most likely being formed in the wind or ejecta. A pre-existing disk would very
likely have been disrupted during the eruption. An alternative model is
self-absorption superposed on a broad emission line. It is easy to invoke
self-absorption and  it probably cannot  be entirely ruled out in this case, but
it is very dependent  on the geometry.  Furthermore, it is  difficult  for
H$\alpha$ to be self-absorbed, and requires that the $n=2$ level be populated in
the intervening material. Given the similarities of the H$\alpha$ and
\ion{Ca}{2} triplet profiles, we would expect them to have the same origin. The
unequal, asymmetric  profiles which change with time, plus the  evidence for 
blue and red-shifted multiple components discussed in \S {6}, in our opinion,
favor an outflow interpretation. 

The double-peaked \ion{Ca}{2} emission lines were first noticed in our spectrum
of 2008 June 20, obtained with our lowest-resolution grating (Gr 13). All three
lines were clearly double, with prominent blue and red components and an easily
recognizable absorption minimum.  The \ion{Ca}{2}  profiles in the echellette 
spectra from 2008 July 6  and  August 30/September 1, shown in Figure~8, are
resolved  and  reveal a  complex and variable structure with evidence for
secondary emission components in  addition to the prominent blue and red peaks. 
The triplet lines are also  asymmetric to the  red, similar to the 
[\ion{Ca}{2}] lines; see \S 5.2. In our RC spectra the triplet lines all appear
single after mid- August 2008.

The first evidence that the strong H$\alpha$ line was double, with a broadened
top and two  maxima of about equal strength,  appears in our 2008 June 21 RC
moderate-resolution spectrum. The H$\alpha$ emission line in the three
echellette spectra from 2008 July/August  shows two well-resolved components
(see Figure~7), with evidence for secondary emission features similar to those
observed in the \ion{Ca}{2} triplet. The H$\alpha$ profile in the last
echellette spectrum, a year later,  from 2009 June 5 still shows a prominent
double-peaked structure with evidence for multiple components. H$\beta$ in the
echellette  spectra  from 2008 also shows similar double profiles. Although the
other hydrogen lines may appear essentially single, inspection of their profiles
shows evidence for broadening due to other emission components. Br$\gamma$ and
Pa$\alpha$ in the low-resolution ISAAC 2~$\mu$m spectrum from 2009 July are in
emission and Br$\gamma$ has a double-peaked profile. 

The series of moderate-resolution RC spectra    reveals some interesting
short-term variability in the  strength of the H$\alpha$ line and the 
double-peaked signature.  For example, in Figure~9, the 2008 September 10
spectrum shows a well-developed double profile  with two clearly  separate
components and an absorption minimum.  In the spectrum from September 13,
however, with the same grating, the separation in  the H$\alpha$ profile was 
barely  detectable as a broadening or small bump on the red side of the line.  
But then only three days later (September 16)  the H$\alpha$ profile was 
clearly double again and the line was even stronger than on September 10. On
September 27, the signature of the split profile was again weaker and the
overall  flux in the H$\alpha$ line  much weaker. After that the hydrogen
emission profiles  were very consistent with little evidence for variability in
the profile shape.  H$\alpha$  continued  to show evidence for the double-peaked
profiles until our last observation  with this grating in 2009 January. These
short-term changes may be indicative of variations in the outflow.  They also
correspond to the change in the slope of the light curve and the SEDs, \S {4},
and the transition to an optically thin state, \S {6}.

The heliocentric Doppler velocities measured from the echellette spectra for
the  components of the  double \ion{Ca}{2} triplet and the H$\alpha$ line are
given in Tables 2 and 3, respectively.    The \ion{Ca}{2}  and  H$\alpha$
profiles show considerable complexity, with evidence for  additional emission
components which appear as shoulders or secondary bumps on the  primary blue and
red peaks.   Comments and notes about these features are included in the
tables.   Because of the complexity of many of these profiles, a program to
automatically fit Gaussians to the profiles did not give satisfactory results.
To measure the velocities of the two components, we therefore first measured the
maximum or  central wavelength of the stronger blue component by bisecting the
very top of the profile. We then fit a Gaussian to the blue component assuming a
symmetric profile about the central wavelength.  This  was then subtracted from
the line to estimate the central wavelength of any secondary features or
shoulders. The same process was followed for the red component. We have also
included the Doppler velocities for the H$\alpha$ blue and red peaks  measured
in the RC spectra.   We followed a process similar to that described above to
measure the velocities of the two primary components and adopted the   
intersection of the curves for the blue and red components for the absorption
minimum. The secondary emission features are not as evident in these
moderate-resolution spectra.

The mean velocities measured for the absorption minimum in the double-peaked
\ion{Ca}{2} and  H$\alpha$ lines from the echellette spectra of $192 \pm 2$ km
s$^{-1}$ and $202 \pm 6$ km s$^{-1}$, respectively,  are consistent  with each
other and  compatible with the  OT's expected radial velocity of $\approx$190 km
s$^{-1}$ from NGC 300's rotation curve at its location  in the galaxy
\citep{Marcelin85}, and with the velocities of other  emission lines in
Table~4. 

\subsection{The [\ion{Ca}{2}] Lines} 

The [\ion{Ca}{2}] lines, formed in a very low-density region,   do not show any
evidence for  double-peaked profiles,  but the profiles of both lines  are
asymmetric to the red and have Thomson   scattering wings extending from 
approximately $\pm$250 to $\pm$300 km s$^{-1}$; see Figure~10.  Berger et al.\
(2009) attributed the asymmetric profiles to inflowing gas on the red side  and
absorption  by intervening material on the blue side of the lines; however,
self-absorption is  highly  unlikely, if not impossible,  for a forbidden
emission line. A physically more realistic explanation  is the effect of
electron scattering on the profiles, as first demonstrated  by Auer \& Van
Blerkom (1972). They showed that radiation transfer through an expanding 
envelope will substantially alter the shape of the profiles.  For a line formed
by recombination, as in the case of the [\ion{Ca}{2}] lines,  the only source of
opacity is the free electrons. With electron-scattering optical depths  of
$\sim$1, the profiles will be asymmetric with extensive wings to the red of the 
expected line center. See  Hillier (1991)  for examples of model profiles with  
the effects of electron scattering which closely resemble the [\ion{Ca}{2}]
lines in the  transient.  SN~2010dn shows similar asymmetric [\ion{Ca}{2}]
profiles (Smith et al.\ 2011). 

After mid-June 2008 (Table 4), the [\ion{Ca}{2}] Doppler velocities measured
from the echellette  spectra and the RC spectra with grating (\#58) show little
evidence for variability,  and yield  mean velocities of  $207 \pm 2.6$ km
s$^{-1}$ and $226 \pm 7.7$ km s$^{-1}$ for the $\lambda$7291~{\AA} and
$\lambda$7323~{\AA} lines, respectively. 

\subsection{Other Emission Lines} 

Many weaker emission lines are present in these spectra. Several are visible  
in the RC spectra, but they are obviously much more numerous in the echellette
spectra.   These lines became relatively stronger at later times. They may have
always been present but become more apparent as the continuum declined with
time,  although most were no longer detectable after early 2008 September.  
Most of these lines are identified with \ion{Fe}{2} and [\ion{Fe}{2}], formed in
the star's wind.   Not surprisingly, emission lines from the Paschen series of
hydrogen are present.  The presence of \ion{He}{1}, \ion{O}{1}, [\ion{O}{1}], 
and \ion{Na}{1} D in emission is of more interest and is discussed here. 

The \ion{O}{1} blend at $\lambda$8446~{\AA} is present in emission in  the  RC
spectra beginning with those obtained in 2008 July,  and in the echellette
spectra, and is still present in the  RC spectra from later times until 2008
December.  In their higher-resolution echelle spectra, \citep{Berger09} find the
\ion{O}{1} $\lambda$8446 line  initially in absorption, which transitions to an
emission line in 2008 July/August. This is similar to the behavior of the
\ion{Na}{1} D lines, described below. Although the $\lambda$8446 line appeared
to increase  in strength with time in our spectra, the total flux in the line
remained  essentially constant. This emission line is observed in novae,
nova-like objects, AGNs, and Seyfert galaxies, and  given the lack of any other
\ion{O}{1} emission lines in the spectrum, its most  likely origin is Ly$\beta$
pumping by fluorescence \citep{Bowen,Grandi80}. Berger et al.\ (2009) also noted
the connection with Ly$\beta$.  This is the only emission line that shows a
blueshift in its velocity with time (Table 4), which may be related to the
fluorescence with Ly$\beta$. The [\ion{O}{1}] lines at $\lambda$6300~{\AA} and
$\lambda$6363~{\AA} are also identified in   two echellette spectra, but are not
seen in later spectra. Berger et al.\ also noted a one-time appearance in their
echelle spectra from 2008 August 23.  

Similarly, the \ion{He}{1} emission line at $\lambda$5876~{\AA} is observed  in
the RC spectra,  beginning with those obtained in 2008 June. Additional,
relatively weak, \ion{He}{1}  lines at  $\lambda\lambda$7065, 6678, 3964
(blended with \ion{Fe}{2}) and $\lambda$3888 (probably blended with H$\zeta$)
are also identified in the echellette  spectra. These weaker \ion{He}{1} lines
are not apparent in the RC spectra. Both the \ion{O}{1} $\lambda$8446 emission
line and the \ion{He}{1} emission lines require UV photons for their excitation.
But, unlike the \ion{O}{1} line, the strong \ion{He}{1} $\lambda$5876 line
slowly weakened with time and was no longer observed in spectra obtained after
2008 October~1. 

The \ion{Na}{1} D lines interestingly change from a strong absorption feature in
the echellette spectrum from 2008 July 6 to emission in the August
30--September~1 spectra shown in Figure~11\footnote{Berger et al.\ (2009)
assumed the \ion{Na}{1} D line absorption lines are interstellar  in origin.
However, strong \ion{Na}{1} absorption is expected in F-type spectra. 
Furthermore the velocities measured for the \ion{Na}{1} absorption lines  are
the same as for the other absorption lines formed in the transient's dense
wind.}.  In the RC spectra, the \ion{Na}{1} D lines are observed in absorption 
at maximum (2008 May 15) and in the spectra from June 20 and July 18. They are
not apparent either in emission or absorption in a spectrum from   August 23, 
but are in emission in the spectrum from  September 8 obtained with the same
grating.  With low resolution it is uncertain if they are present in emission
after that.  Thus in about one month, $\sim$100 days after maximum,  the
\ion{Na}{1} D lines went from absorption to emission.   The  line profile from
August 30--September~1 shows some weak absorption superposed on the emission
(Figure~11) . Measurement of the total flux in the absorption components show
that they decreased by about a factor of two between  July 6 and  August 30. So,
about half the flux previously in the absorption lines was  now in emission. 
The \ion{O}{1} $\lambda$8446 line also transitioned from absorption to emission
during the same time period (Berger et al.\ 2009). Since the other absorption
lines were also  weakening with time, we  suggest that {\it the expanding
envelope was becoming more transparent to the ionizing  photons.}  

The velocities measured from these different lines are summarized in Table 4. 
All velocities are heliocentric. The mean velocity of 40 emission lines in the 
2008 August 30/September~1 echelette spectra is $+200 \pm 4.4$ km s$^{-1}$. 

\subsection{The Absorption Spectrum}

Several absorption lines typical of F-type supergiants and optically thick
winds   are  present in the spectrum at maximum light,  although they appear to
weaken with  time in the low-resolution RC spectra. These lines include several
luminosity-sensitive  features such as the \ion{O}{1} triplet at
$\lambda$7774~{\AA}, and the \ion{Fe}{2}--\ion{Ti}{2} blends at
$\lambda$4172-4178~{\AA},  plus \ion{Ca}{2} H and K, Ca I $\lambda$ 4226~{\AA},
the \ion{Na}{1} D lines, and a weak G-band. These same lines, plus additional
lines of \ion{Sr}{2}, \ion{Ti}{2}, \ion{Sc}{2}, and \ion{Mn}{1}, appear
relatively strong in the echellette spectrum from 2008 July~6, but at
significantly lower velocities (Table 4), and are no longer  apparent in the 
August 30--September~1 spectra. Their  mean radial velocity in the July~6
spectrum  from 33 absorption lines,  $180 \pm 1.7$ km s$^{-1}$, is somewhat less
than  the velocities of the  absorption minima in the double-peaked profiles and
most  of the emission lines measured at later times.  

Very interestingly, in the echellette spectrum from 2008 July~6, the \ion{Ca}{2}
H and K absorption lines are apparently double, with  two absorption minima
(Figure~12). Evidence  for the two absorption minima can still be seen in the
spectrum from August 30, although the K line is much weaker.  The red-shifted
component has a velocity similar to that measured for the H and K lines at
maximum light, while the deeper minimum has a velocity like that of the other
absorption lines and presumably originates in the same material.  Although the
signal-to-noise is low, evidence for the same double absorption can be seen  in
the profiles published by Berger et al.\ (2009), although  they attributed the
width and shape of the H and K line profiles to infall and outflow in the ejecta
by analogy with a  model for the ejecta for IRC~+10420 (Humphreys et al.\ 2002).
The arguments in favor of both infall and outflow for IRC~+10420 however were
based on the presence of inverse P Cygni profiles primarily in emission lines of
\ion{Fe}{2}, some of which showed both P Cygni and inverse P Cygni profiles.
Neither are observed in the NGC 300 transient. 

As mentioned in \S {5.3}, the  \ion{Na}{1} D absorption lines transitioned from 
absorption to emission in about 30~days.  The \ion{O}{1} triplet at
$\lambda$7774~{\AA} is a well-known luminosity indicator for  intermediate-type
supergiants.    Its velocities agree with those of the other absorption lines
measured at the same time. It thus originates in the same material, the
transient's wind or envelope \footnote{The \ion{O}{1} absorption line as well as
the absorption minima in the double-peaked \ion{Ca}{2} triplet lines are not
interstellar as suggested by Berger et al. While interstellar absorption is
frequently observed in the resonance transitions of \ion{Ca}{2} (H and K lines)
and \ion{Na}{1} D, the \ion{O}{1} and the \ion{Ca}{2} triplet lines are not
resonance lines.}.  Its mean equivalent width, measured from six spectra from
July~6 to August~24, is  $1.62 \pm 0.08$~{\AA}, corresponding to an $M_V$ of
$-6.8 \pm 0.2$  (Osmer 1972), which coincidentally is consistent with  $M_V$ for
the pre-eruption object, assuming no bolometric correction. 

We have also tentatively identified \ion{Ba}{2} in absorption at 4554~{\AA} and
4934~{\AA} in  the echellette spectrum from 2008 July~6. Their respective
equivalent widths are 0.5 and 0.4~{\AA}, with velocities of 220 and 197 km
s$^{-1}$, slightly higher than the velocities of the other absorption lines in
Table 4. Barium is an $s$-process element, so if this identification is correct,
these lines  may have  important implications regarding  the evolutionary state
of the star discussed in the last section.  The \ion{Ba}{2} lines, however, are
also stronger in stars of  high luminosity. In the absence of an abundance
analysis, these lines could be enhanced in the low-gravity environment of the
wind and not necessarily be due to the $s$-process in the interior of an AGB
star.  

As already noted, the absorption lines weaken with time. The spectrum
transitions from one with numerous absorption lines in our 2008 July echellette
spectrum to one dominated by emission lines by  2008  September and later, 
probably  due to the expansion and decreasing density of the envelope. To
further illustrate this transition, we show these   echellette spectra in the
blue in  Figure 13.  None of the spectra obtained after early 2008 September  
show any evidence for absorption lines. 

\section{The Transient's Post-Eruption Outflow and Wind}

The spectra of the NGC 300 OT show an object whose outflow and wind were in
transition  for the first 30+ days after the eruption or maximum light observed
on 2008 May 15. The outflow pattern continued to develop and change for the next
$\sim$100 days, but after that, from approximately  mid-September 2008 to our
 spectrum from  2009 January, the  spectra are very similar and are dominated
by the emission-line spectrum in the wind or ejecta. The transient thus provides
an opportunity to observe a post-eruption and recovery in progress, but
unfortunately, moderate- to high-resolution spectra with good signal-to-noise
are lacking during the first three weeks or so. Consequently, in this section,
we  concentrate our discussion on the transient's  spectrum from approximately
30 days after the observed maximum. 

\subsection{The Eruption} 

Our first spectrum, from 2008 May 15, corresponds  to the observed maximum light
(Bond et al.\ 2009). The hydrogen, \ion{Ca}{2}, and [\ion{Ca}{2}] lines and the
\ion{Ca}{2} H and K absorption lines at that time have Doppler velocities of
400--500 km s$^{-1}$ and  are significantly red-shifted with respect to their
velocities measured at later times (Table 4).  None of the  emission lines show
evidence for the double-peaked profiles, which only appeared later. The
\ion{Ca}{2} triplet, however, has a very complex, broadened, and red-shifted
profile, with  multiple peaks perhaps indicative of multiple shells moving
outwards.  Berger et al.'s (2009) low-resolution  spectrum from 2008 May 23
shows similar features in the \ion{Ca}{2} triplet complex consistent with this
description. The absorption lines also have higher measured Doppler velocities
comparable to the emission lines (Table 4).  Approximately 30 days later,
mid-June 2008, the measured velocities are significantly  lower, $\sim$250 km
s$^{-1}$, and consistent with those published by Berger et al.\ from the same
period. High-resolution spectra published by Berger et al.\ show complex
\ion{Ca}{2} triplet profiles with two to three possible  red-shifted components,
which may be due to absorption in intervening material and infall as Berger et
al.\ suggest. However,  the deep absorption  minimum in the \ion{Ca}{2} lines at
zero velocity in their Figure 14 is not interstellar. 

\subsection{Post-Eruption} 

The transition to lower velocities observed for both the emission and absorption
lines is perhaps best explained as an initial eruption, followed, at 
approximately 30 days, by the appearance of a dense wind or envelope responsible
for the absorption lines, and the possible development of a bipolar outflow
discussed below. Residual evidence for the expanding gas from the initial
eruption is observed  in the red-shifted absorption component of the \ion{Ca}{2}
H and K lines and possibly in the  secondary emission features and broadened 
profiles of  H$\alpha$ and \ion{Ca}{2} triplet.  The [\ion{Ca}{2}] lines and the
other emission lines  that appear at later times must originate in the
low-density gas in the surrounding ejecta. As mentioned earlier, the expected
velocity of the transient is $\sim$190 km s$^{-1}$, from  the NGC~300 galactic
rotation curve. Based on the velocities of the emission lines, especially  those
that appear after 30 days, and the absorption minima in the double-peaked 
profiles, we suggest that the transient's systemic velocity is  $\approx$200 km
s$^{-1}$. The absorption lines with a lower mean velocity of $\approx$180 km
s$^{-1}$ originate in the dense slowly expanding wind.

As explained earlier, the double-peaked profiles may be attributed to a bipolar
outflow, a rotating disk, or to self-absorption. We favor the outflow
interpretation, for the  reasons given in \S 5.1. Furthermore, the blue- and
red-shifted primary and secondary emission  components have relatively symmetric
velocities, supporting an  outflow model. It is not necessary to invoke a
complex pattern of infall and outflow to explain the hydrogen and \ion{Ca}{2}
emission profiles observed at later times. 

We  use two approaches to estimate the expansion velocity. When  the absorption
minimum  between the blue and red emission components is clearly  present, the
expansion velocities of the two components is simply the  difference between
their peak velocities and that of the absorption minimum.  When there is no
well-defined minimum, the estimated expansion velocity is simply half the
difference  in the velocities of the two maxima.  The resulting expansion
velocities from both  methods are summarized in Table 5.  

The average  velocities from both the \ion{Ca}{2} and hydrogen emission  lines
in the echelle spectra   indicate  expansion velocities for the primary wind of 
$\sim$70--80  km s$^{-1}$ with respect to the star,   with a range from  a
possible high of 90 km s$^{-1}$ for the blue component and  a low of 65 km
s$^{-1}$ for the red feature.  Evidence that the blue and red features may have
slightly different expansion velocities comes primarily from the measurements 
of the \ion{Ca}{2}  triplet.   The RC spectra yield slightly higher expansion
velocities for the H$\alpha$ lines,  on the order of  100 km s$^{-1}$. This
result is not surprising at this lower resolution, given that the emission lines
may be  broadened by  contributions from faster-moving gas. The measurements
from the RC spectra also  show some  evidence for a higher velocity for the
blue-shifted feature.   Faster-moving gas is also present in the ejecta, as
indicated by the width of the lines, and by the frequent presence of measurable
secondary emission components, or  shoulders, on the line  profiles and
secondary peaks in a few cases. This secondary, faster wind may also be bipolar
and  moving at $\approx$ 160 km s$^{-1}$ relative to the star.  

We also note that in  both the \ion{Ca}{2} lines and H$\alpha$,  the blue
component  slowly increases in strength relative to the red component. This
observation supports our conclusion that the outflow and eruption are  bipolar,
with the nearer, blue-shifted lobe expanding towards us and increasing in
relative intensity with respect to the  receding feature. The $b/r$ ratio is
also included in Table 5. 

An outflow velocity of $\sim$75 km s$^{-1}$  is typical of F-type supergiants, 
and is somewhat lower than the winds associated with LBVs in eruption. It is
similar to  the   expansion velocity, $\sim$60 km s$^{-1}$, for the
well-studied  post-red supergiant IRC~+10420 (Jones et al.\ 1993; Humphreys et
al.\ 2002),   measured from its double-peaked hydrogen and \ion{Ca}{2} emission
lines.   Although  we do not know where the emitting gas is located with respect
to the star,   we assume that the much stronger primary blue and  red emission
components originate in   material closer to the star and that   the  outflow
at  75 km s$^{-1}$ is therefore  representative of the star  in its 
post-eruption  state.  The faster-moving gas  at $\pm$ $\sim$ 160 km s$^{-1}$ 
may have been  produced in the initial eruption. The Doppler velocity of its 
red-shifted component  is comparable to the higher  velocities measured at
maximum light.   

Our model for the NGC 300 OT in post-eruption is a bipolar outflow, perhaps 
surrounded by low-density ejecta, with a slowly expanding  dense wind or
envelope  producing the absorption-line spectrum, which at about 100 days after
the eruption  became transparent to the  ionizing radiation. At later times +120
days, we see only the emission-line  spectrum from the ejecta.  The ionizing
radiation and the increase in the UV flux could be due to the presence of a hot
companion, which came apparent at later times as the dense wind expanded, or to
the underlying hotter layers of the erupting star which had expelled its outer
envelope. 

\section{On the Nature and Evolutionary State of the 2008 Optical Transient in  NGC 300} 

With its strong, narrow hydrogen emission lines,  \ion{Ca}{2} and [\ion{Ca}{2}]
emission, and F-type supergiant absorption spectrum, the OT 
spectroscopically resembled the  post-red supergiant IRC~+10420 (Jones et al.\
1993; Humphreys et al.\ 2002) and  Var~A in M33 \citep{RMH06}, although these
two hypergiant stars are much more luminous.  Their spectroscopic similarities
are telling us that the physical  conditions in these stars' winds and
circumstellar  ejecta are similar, not necessarily that they are the same kind
of star, but like the transient, IRC~+10420 and Var A have also recently 
experienced high-mass-loss  episodes. 

In other respects, the NGC 300 transient is  most like SN~2008S \citep{Prieto08A},  a
``supernova impostor'' (Smith et al.\ 2008).  SN~2008S was also heavily 
enshrouded by dust prior to outburst, with an intrinsic luminosity of  $3.5
\times 10^4 \, L_{\odot}$ from its infrared SED,  implying  an initial mass 
near $\sim$$10\,M_{\odot}$ if it was a red supergiant.  It reached a visual
luminosity of $M_{V}\simeq-14$ at maximum, increasing $\sim$1000 times in
luminosity during the eruption. Like the 2008 NGC 300 transient, an early
spectrum showed  \ion{Ca}{2} and the  rare [\ion{Ca}{2}] lines in emission
\citep{Steele08}. SN~2010da likewise had a heavily obscured progenitor (Khan et
al.\ 2010a; Berger \&  Chornock 2010). However, it is bluer and its  spectrum
showed significant differences. There was apparently  no [\ion{Ca}{2}] emission,
but \ion{He}{2} $\lambda$4686~{\AA} has been identified (Elias-Rosa et al.\
2010; Chornock \&  Berger 2010). \ion{He}{2} emission requires a very hot source
of radiation producing more than 54 eV\null. Binder et al.\ (2011) have detected
an X-ray source identified with SN~2010da and suggest  that it is a supergiant
X-ray binary with a compact companion. SN~2010dn was spectroscopically very
similar to the NGC 300 transient and SN~2008S (Smith et al.\ 2011). Although 
there is only an upper limit to its pre-outburst infrared flux (Berger 2010),
its increase in  apparent brightness by at least 7 magnitudes at maximum,
suggest that the progenitor was very likely obscured. The progenitors of
PTF~10fqs and PTF~10acbp were also not detected in the infrared, but  are
spectroscopically similar to these transients at maximum light (Kasliwal et al.\
2010, 2011), and likewise had a large increase in apparent brightness. 

The positions of the progenitors of the three transients with pre-eruption
detections in the infrared are shown on an HR diagram in Figure 14. Known
LBVs/S~Doradus variables and  several warm and cool hypergiants known for their
instabilities and high-mass-loss  episodes are also shown for comparison.  

With the large amount of obscuring dust, the NGC 300 OT, SN~2008S, and
SN~2010da  must have experienced high mass-loss rates as  post-main-sequence
stars. At  the relatively  high luminosities of their  progenitors,  they were 
most likely red supergiants (RSGs),  stars at the tip of  the AGB, or
post-AGB/RSG stars. The luminosities of the progenitors of the NGC 300 transient
and SN~2008S imply relatively high initial masses, $\geq$$10\,M_{\odot}$, but
they are both also close to the AGB limit.  SN~2008S is just below it, and 
could have reached its  luminosity  as a somewhat lower-mass star on the
AGB\null. Thompson et al.\ (2009) and  Botticella et al.\ (2009) have suggested
that these  stars may represent a new class of less-luminous 
SNe due to electron-capture  with progenitors in the mass range 8--$11\,M_{\odot}$. By comparison with
multi-epoch  {\it Spitzer\/} IRAC observations in M33, Thompson et al.\
identified a group of  reddened stars with similar colors that they call
extreme-AGB stars that may be the progenitor class, and Khan et al.\ (2010b)
have identified a limited population of similar self-obscured stars in four
nearby galaxies, including M33, suggesting that this is a short-lived high mass
loss phase.  

Here we suggest that the progenitor of the NGC 300 transient, and by analogy,
SN~2008S and others described above,  was an optically obscured evolved star
possibly related to the OH/IR stars observed in our galaxy and the Magellanic
Clouds.   Given their lack of any obvious variability from multiple  IRAC
observations (Thompson et al.\ 2009), the NGC 300 transient and SN~2008S were no
longer   fundamental-mode pulsators, i.e., they were not  Mira variables.   Thus
they had  very likely left the region of  the RSGs and AGB stars on the HR
diagram and were on a blue-loop back to warmer  temperatures. For that reason
all three transients in Figure 14 are plotted as horizontal lines extending
from the RSG/AGB region to intermediate temperatures.  

Given the large amount of mass the  transient  must have lost as an  RSG
or AGB star, implied by the heavy obscuration, it must  be highly evolved, 
possibly on a second blue loop. This suggestion is   supported by the
carbon-based mid-IR  emission feature (Prieto et al.\ 2009)  observed in
proto-planetary nebulae and also the \ion{Ba}{2} lines if they are due to the
$s$-process. The emission feature  implies carbon-rich ejecta and stars with the
\ion{Ba}{2} 4554~{\AA} line typically show enhanced carbon features (CN, CH, and
C$_{2}$). But there are no carbon features in the spectra of the wind or
envelope, and instead the spectrum has strong \ion{O}{1} absorption and emission
lines.  

The luminosity of the NGC 300 OT progenitor places it slightly above the AGB
limit and  implies an initial  mass of 10--$15\,M_{\odot}$ if it was a true
supergiant, and  Gogarten et al.\ (2009)  estimated a mass range of 13--$17\,
M_{\odot}$ based on the CMD of the associated stellar  population. Our
inspection of the transient's immediate environment was inconclusive for the 
mass of the progenitor. The mid-IR emission feature, however,  supports a
post-AGB origin for the progenitor and by implication a lower mass,
$<$$10\,M_{\odot}$. These  results are not necessarily inconsistent. Stars 
with  initial masses in the 5--$9 \,M_{\odot}$  range, when at the tip of the
AGB,  can reach  the luminosities of the NGC~300 OT and SN~2008S progenitors.
Thus both objects could have originated from stars with initial masses in this
range,  for which the later stages of evolution are  somewhat uncertain.
Furthermore there is no reason why a true supergiant could not pass through the
RSG region more than once. Evolutionary tracks for stars in the 9--$12\,
M_{\odot}$ range show a second approach to the RSG region \cite{MM00}.

The presence of \ion{Ca}{2} and [\ion{Ca}{2}] emission, observed in the winds of
the  F-type hypergiants, also suggests that the progenitor was  not as hot as a
B-type supergiant, as in the case of LBV eruptions, but may have been  somewhat
cooler, more likely an intermediate-temperature star. The implication that  the
underlying star was also  somewhat warmer than an RSG or AGB star is supported
by  its  bipolar  outflow  with  speeds of $\sim$  75 -- 100  km s$^{-1}$ near 
the star and  somewhat  faster-moving ejecta   at $\approx$ 160  km s$^{-1}$,
probably further from the star.  These velocities are  significantly  higher 
than the winds associated with RSGs, AGB stars, and the OH/IR stars
($\sim$15--25 km s$^{-1}$ for the AGB stars, and up to 40 km s$^{-1}$ for the
OH/IR supergiants).  The [\ion{Ca}{2}] emission, however, originates in
slower-moving, lower-density gas, possibly  a remnant from the progenitor's
earlier state, as an AGB star or RSG.

The transient  was thus most likely an evolved intermediate-mass  star in
transition to warmer temperatures that had recently experienced high mass
loss.    Having already shed a lot of mass, these stars are  at or
near the Eddington-limit for their luminosities. In their post-RSG or post-AGB 
evolution, they will encounter a temperature regime (6000--9000~K) of increased
dynamical instability  \citep{deJ98}  driven by the  increasing opacity in
combination with increasing rotation and pulsation.  For example,  the 
less-luminous LBVs \citep{HD94} are generally considered to be post-red
supergiants.   With their reduced mass  they are much closer to the Eddington
limit for their luminosities and  are therefore  more subject to radiation
pressure and other  instabilities.   Although it is not known if these
instabilities can lead to a major eruption, the most famous  example may be the
progenitor of SN~1987A\null. Its  interlocking  bipolar loops and equatorial
ring are evidence for high-mass-loss  events prior to the terminal explosion.
Although the rings have been attributed  to the RSG stage, they closely resemble
the ejecta around many LBVs \citep{HD94},  and Smith (2007) has  emphasized
their resemblance to the ejecta associated with the LBV  candidate HD~168625. 
However, the transient's bipolar eruption may be an equally likely model for 
SN~1987A's pre-supernova ejecta and for high-mass-loss episodes for post-RSG
stars evolving to the LBV stage as well as post-AGBs/proto-planetary nebulae (see Soker \& Kashi 
 2011).  
In this case, the onset of the ionizing radiation and increased UV flux required
for the  observed transition to the emission line spectrum may be due to an
underlying hotter zone, revealed when the expanded cool, dense wind or false 
photosphere dissipated. 

The NGC 300 2008 OT and related objects like SN~2008S are not classical or  normal
LBVs\slash S~Dor variables. Although they have  experienced a ``giant
eruption,'' increasing their total luminosity by 100 to 1000 times, they are
also not  examples of giant-eruption  LBVs \citep{HD94}. The progenitors are apparently less
luminous,  intermediate-mass  stars on a blue loop that have previously shed a
lot of  mass.    Although the origin of the NGC 300 transient's  eruption is not
known, it briefly increased its total luminosity by about 250 times and 
released $10^{47}$ ergs. The eruption produced a cool dense envelope  and a
two-component bipolar outflow. It may have been some type of yet unexplained
failed supernova  or a sub-photospheric eruption expelling up to $1\,M_{\odot}$
in transit to a warmer state.

\acknowledgments

The authors gratefully acknowledge the support of the SMARTS consortium to
obtain the  multi-wavelength photometry and the RC series of spectra. We thank the
European Southern  Observatory for Director's Discretionary Time for the
near-infrared spectra in 2009 July,  VLT/UT1/ISAAC proposal:
283.D-5019(A,B,C,D).  We also thank Elena Sabbi for her assistance with analysis of
the CMD and determining the TRGB distance to NGC~300,  and    G. Pietrzy\'nski
and Julio Chanam\'e who obtained the MagE spectra on 2008 July~6 and 2009 June~5,
respectively. Prieto acknowledges support from NASA through Hubble
Fellowship grant HF-51261.01-A awarded by STScI, which is operated
by AURA, Inc. for NASA, under contract NAS 5-2655.

\appendix{{\bf Appendix} Table A1 and A2} 
Table A1 (Photometry) and Table A2 Journal of Spectroscopic Observations will be in the electronic version.


\begin{deluxetable}{ccc}
\tablenum{1} 
\tablecaption{RC Spectrograph Gratings}
\tablewidth{0pt}
\tablehead{
\colhead{Grating} &  \colhead{Wavelength Range} & \colhead{Resolution} \\ 
 &    \colhead{{\AA}}  &  \colhead{{\AA}}
 }
\startdata
9   &   4770-7180 &  8.6 \\
13   &   3150-9350 &   17.2 \\
26   &  3660-5440 &   4.3  \\
47   &   5650-6970 &  3.1 \\ 
58  &    6000-9000 &  6.5 \\
\enddata
\end{deluxetable}

\begin{deluxetable}{llrrrl}
\tabletypesize{\footnotesize}
\tablenum{2} 
\tablecaption{Measured Velocities for the Ca II Triplet Emission Lines from the Echellette Spectra}
\tablewidth{0pt}
\tablehead{
\colhead{U.T. Date} & \colhead{Line Id.} & \colhead{Blue Pk.} & \colhead{Abs. Min.} & 
\colhead{Red Pk.} &  \colhead{Comment} \\
\colhead{2008} &  \colhead{{\AA}} &  \colhead{km s$^{-1}$} & \colhead{km s$^{-1}$} & \colhead{km s$^{-1}$} &     
}
\startdata
Jul 06  &  $\lambda$8498 primary  &   120  &  198   & 266  &  \nodata \\
        &    \nodata  secondary   &    10  &  \nodata & 336 & \tablenotemark{a} \\ 
        &  $\lambda$8542   &    67  & 189  & 304   & asymmetric to red \tablenotemark{b}\\
        &  $\lambda$8662 primary  &  97   & 190  & 294  &  blue \& red peaks appear double \tablenotemark{c} \\ 
        &    \nodata  secondary   &  61   &  \nodata & 378  & \nodata \\ 
Aug 30 &  $\lambda$8498    &    117  & \nodata &  265\tablenotemark{d} & line appears single \tablenotemark{d} \\
       &  $\lambda$8542 primary   &     107  &  188  & 251  &  \nodata \\
       &    \nodata  secondary    &      35  & \nodata & 400 & \tablenotemark{e} \\  
       &  $\lambda$8662 primary   &     105  &  201  &  254  & \nodata \\
       &    \nodata  secondary   &      49   & \nodata & \nodata  &  \tablenotemark{f} \\
Sep 01  &  $\lambda$8498   &  123  &  \nodata  &  248 &  \tablenotemark{g} \\
         & $\lambda$8542 primary   &  96 &  193  & 242  &  \nodata \\
         &    \nodata  secondary    & 31 &  \nodata & 349 & \tablenotemark{h} \\
         & $\lambda$8662  primary &  119  & 183   & 249  &   \nodata \\
         & \nodata  secondary     &   69  & \nodata & 383 & \tablenotemark{e}\\
\enddata
\tabletypesize{\footnotesize}
\tablenotetext{a}{An emission shoulder is present on the blue and red sides, respectively of the two emission peaks.} 
\tablenotetext{b}{The red component is asymmetric with a wing extending to 8570 {\AA}.}
\tablenotetext{c}{The blue and red-shifted components each have two approximately equal peaks. The red component also has a prominent redward extending wing.} 
\tablenotetext{d}{The line appears single with a small shoulder on the red side. There is no absorption minimum.}
\tablenotetext{e}{The blue side of the profile has a small secondary peak and the red component has a red shoulder.}  
\tablenotetext{f}{The blue side of the profile has a small secondary peak. There is no shoulder on the red component.} 
\tablenotetext{g}{The line appears single with a prominent shoulder on the red side and small shoulder or bump on the blue side with a velocity of -8 km s$^{-1}$.}
\tablenotetext{h}{The blue side of the profile has a small secondary peak, and the red side has a shoulder.}

\end{deluxetable}

\begin{deluxetable}{llrrrl}
\tabletypesize{\footnotesize}
\tablenum{3} 
\tablecaption{Measured Velocities for the Double--Peaked  Hydrogen Emission Lines }
\tablewidth{0pt}
\tablehead{
\colhead{U.T. Date}  & \colhead{Line Id.} & \colhead{Blue Pk.} & \colhead{Abs. Min
.} & \colhead{Red Pk.} &  \colhead{Comment} \\
 &     &  \colhead{km s$^{-1}$} & \colhead{km s$^{-1}$} & \colhead{km s$^{-1}$} &   }
\startdata
Echellette and ISAAC Spectra &  &  &  &   &  \\ 
2008 Jul 06    &   H$\alpha$ primary  &  141 & 215 &  297 & \nodata  \\ 
        &   \nodata  secondary &  18 & \nodata & 433 &  \tablenotemark{a} \\ 
        &   H$\beta$           &  128 & \nodata & 370 & \tablenotemark{b} \\ 
2008 Aug 30  &    H$\alpha$ primary  &  137  &  212 &  265 & \tablenotemark{c} \\
        &    \nodata  secondary &  \nodata & \nodata & 446 & red shoulder \\ 
        &     H$\beta$          &  113      &  207   &  262    &  \nodata \\ 
2008 Sep 01  &  H$\alpha$ primary   &  126      &  192    &  258    &  \nodata  \\
        & H$\beta$           &    156      & 236     &  283    &  \nodata \\
2009 Jun 05   &  H$\alpha$  primary & 112   & 188   &  239     & \nodata \\  
        &   \nodata  secondary &  24 &  \nodata &  \nodata &   \nodata  \\
2009 Jul 19 -- 23 & Br$\gamma$      & 86   &  184     &  258    & VLT/ISAAC \\ 

H$\alpha$ -- RC Spectra with grating \#47, all 2008 &  &  &  &  &  \\ 
Jun 21 &   \nodata            &  107   &  205  &  286  &   \tablenotemark{d} \\
Aug 25 &    \nodata           & 78    &  216  &  293  &  \nodata \\
Sep 10 &   \nodata            &   94    &  207  &  293  &  \nodata \\
Sep 13 &    \nodata            &  90    & \nodata & 318 &   \tablenotemark{e} \\
Sep 16 &    \nodata            &  107   & 241:  &  301  &  \nodata \\
Sep 27 &    \nodata            &  112   & \nodata & 323 &   \tablenotemark{e} \\
Oct 13 &     \nodata            &  104   & \nodata & 358 &   \tablenotemark{e} \\
Nov 03 &    \nodata           &  112   & \nodata & 279 &   \tablenotemark{e} \\
Nov 09 &     \nodata           &  140   & \nodata & 282 &   \tablenotemark{e} \\
\enddata
\tablenotetext{a}{The blue and red peaks each have a small shoulder.}   
\tablenotetext{b}{Red shoulder; no absorption minimum.} 
\tablenotetext{c}{The blue-shifted emission profile is very broad, this velocity is for the 
peak.}
\tablenotetext{d}{Two small maxima at the top of an otherwise single profile.}
\tablenotetext{e}{No measurable absorption minimum}
\end{deluxetable}

\begin{deluxetable}{llll}
\tabletypesize{\footnotesize}
\tablenum{4} 
\tablecaption{Radial Velocity Summary for the Emission and Absorption Lines}
\tablewidth{0pt}
\tablehead{
\colhead{Identification}  &  \colhead{U.T. Date} &  \colhead{Velocity} & \colhead{Comment} \\
&   \colhead{2008}   &    \colhead{km s$^{-1}$}  &  
}
\startdata
Emission Lines &  &   &   \\  
H$\alpha$      &       15 May    &    395     &    FWHM  1050 km s$^{-1}$, Gr \#13\\ 
               &       20 Jun    &    201     &    FWHM   981 km s$^{-1}$, Gr \#13\\    
H (Paschen lines) &  30 Aug -- 01 Sep & 155 $\pm$ 8 &  6 lines, echellette\\   
{[Ca II]} ($\lambda\lambda$7291,7323)  &  15 May    &   478, 363 &    Gr \# 13 \\
                                     &  20 Jun    &   250, 238 &    Gr \# 13 \\
                                     &  06 Jul    &   208, 229 &  echellette \\
				     &  30 Aug -- 01 Sep    &   202, 222 &  echellette \\
				     & 06 Jul -- 19 Dec & 208, 227 & mean, 7 obs., Gr \#58\\
He I ($\lambda$5876)                 &  20 Jun    &   147:      &  Gr \#13\\  
                                     &  06 Jul    &   155      &  echellette\\
				     &  30 Aug -- 01 Sep  &  155   &  echellette\\
O I ($\lambda$8446)                  &  06 Jul    &   215      &  echellette\\
                                     &  07 Aug    &   218      &  Gr \#58\\
				     &  24 Aug    &   126      &  Gr \#58\\
				     &  30 Aug -- 01 Sep  &   153      &  echellette\\
				     &  21 Sep    &   108      &  Gr \#58\\
				     &  07 Nov    &   109      &  Gr \#58\\ 
{[O I]} ($\lambda\lambda$6300,6363)    &  06 Jul    &  197,---   &  echellette\\
                                     &  30 Aug -- 01 Sep   &  205,204  &  echellette\\
Na I D ($\lambda\lambda$5890,5896)   &  24 Aug    &  210,245   &  Gr \#47\\  
                                     & 30Aug-01Sep & 252,255   &  echellette\\
Fe II ( 7 lines)                     &  06 Jul    &  230 $\pm$ 10 & echellette\\
  ''      ( 12 lines)                & 30Aug-01Sep & 210 $\pm$ 4  & echellette\\  
{[Fe II]} ( 2 lines)                 & 06 jul     &  205 $\pm$ 3  & echellette\\
  ''        ( 14 lines)              & 30Aug-01Sep &  207 $\pm$ 6 & echellette\\

Absorption Lines  &    &  &   \\   
Ca II H             &  15 May  &    409    & Gr \#13\\
                    &  06 Jul  & 191,342   & echellette, double abs.\\
		    &  30 Aug -- 01 Sep  &  182       & echellette\\
Ca II K             &  15 May  &    453    & Gr \#13\\
                    &  06 Jul  & 181,402   & echellette, double abs.\\
		    &  30 Aug -- 01 Sep  & 195,432   & echellette, double abs.\\
Na I D (5890,5896)  &  20 Jun  & 184       & blend, \#13\\
                    &  06 Jul  & 176,175   & echellette\\
		    &  30 Aug -- 01 Sep  & 177,184   & echellette\\
O I ($\lambda$7774)  & 06 Jul  &  140:    &  Gr \#58         \\   
                     & 06 Jul  &  183    &  echellette\\
		     & 01 Aug  &  208    &  Gr \#58         \\
		     & 07 Aug  &  218    &  Gr \#58         \\
		     & 30 Aug -- 01 Sep  &  178  & echellette\\ 
Ba II: ($\lambda$4554,4934) & 06 Jul  & 220, 197 & echellette\\ 
Other Absorption lines &  06 Jul    &  179 $\pm$ 2.2    & echellette, 24 lines  \\
 (Sr II, Ca I, Mn I, Ti II, V II)  &        &           &              \\

\enddata
\end{deluxetable}

\begin{deluxetable}{llcccc}
\tabletypesize{\footnotesize}
\tablecaption{Expansion Velocities from the Double Ca II and Hydrogen Emission Lines}
\tablenum{5}
\tablewidth{0pt}
\tablehead{
\colhead{U.T. Date} & \colhead{Line Id.({\AA})} & \colhead{Blue Pk.}  &   \colhead{Red Pk.}
&  \colhead{Peak to Peak}  & b/r \\
& & \colhead{km s$^{-1}$}    &   \colhead{km s$^{-1}$} &    \colhead{km s$^{-1}$}  &    
} 
\startdata
Ca II Triplet  &   &   &   &   &  \\ 
2008 Jul 06  &  $\lambda$8498 primary &  -78  &  68  &  73  &  1.13 \\
         &  \nodata secondary     &  \nodata & \nodata & 163 &  \nodata\\
         &  $\lambda$8542 &  -122  &  115  &  118  & 1.17 \\
         &  $\lambda$8662 primary &  -93  &  104  &  99 &  1.27 \\
         &  \nodata secondary     & nodata & \nodata &  160 &  \nodata \\
2008 Aug 30  &  $\lambda$8498 &  \nodata & \nodata &  74  &  \nodata \\
         &  $\lambda$8542 primary  &  -81  &  63  &  72  &  1.40 \\
         &  \nodata secondary      &  \nodata & \nodata &  183 & \nodata \\
         &  $\lambda$8662 primary & -96  &  53  &    75  &  1.46 \\
2008 Sep 01   &  $\lambda$8498 &  \nodata & \nodata &  63 &  \nodata \\ 
         &  $\lambda$8542 primary  &   -97 &  49  &  73  & 1.38 \\
         &  \nodata secondary      &  \nodata & \nodata &  159 & \nodata \\
         &  $\lambda$8662 primary & -64  &  66 &  65  & 1.47 \\ 
         &  \nodata secondary      &  \nodata    & \nodata  & 157 & \nodata   \\ 
Avg      &   primary       &  -90 $\pm$ 6.4  & 74 $\pm$ 8.9  & 79 $\pm$ 5.6  & \\
         &   secondary     &   \nodata & \nodata &  164 $\pm$ 4.2  & \\
Hydrogen -- Echellette and ISAAC  &   &  &  &  &   \\   
2008 Jul 06  &  H$\alpha$ primary & -74  &  72 &  78 &  1.06 \\
         &  \nodata secondary &  \nodata & \nodata & 207 &  \nodata\\
         &  H$\beta$          &  \nodata & \nodata &  121 &  1.38 \\
2008 Aug 30  &  H$\alpha$ primary &  -75  &  53  &  64  &  1.32 \\
         &   H$\beta$          & -94  &  55  &  75   &  1.88 \\
2008 Sep 01  &  H$\alpha$ primary   &  -66  &  66  &  66   &  1.36  \\
         &   H$\beta$          &  -80  &  47  & 64  & 1.83 \\
2009 Jun 05   &  H$\alpha$ primary   &  -76  &  51    &  64  &  1.5 \\ 
2009 Jul 19-23 & Br$\gamma$          &  -93     &  74      & 86     & \nodata     \\ 

Avg     &  H$\alpha$ primary  &  -73 $\pm$ 2.0 &  60 $\pm$ 4.7 & 68 $\pm$ 2.9 &  \\
        &   H$\beta$          &   -87 $\pm$ 7 & 51 $\pm$ 4 & 70 $\pm$ 5 &  \\ 
H$\alpha$ -- RC Spectra (grating \#47), all 2008 &   &   &   &   &  \\  
Jun 21  &  \nodata        &  -98  & 81 &  90  & 1.03 \\
Aug 25  &   \nodata       &  -138  & 77 &  108 &  \nodata \\
Sep 10  &    \nodata      &  -113  & 86 &  100  &  1.24 \\
Sep 13  &    \nodata      &   \nodata & \nodata & 114  & \nodata \\
Sep 16  &   \nodata      &   -134:  & 60: &  97  &  1.37 \\
Sep 27  &   \nodata      &   \nodata  & \nodata &  106  & \nodata \\ 
Oct 13  &   \nodata      &   \nodata  & \nodata &  127 &  1.42 \\
Nov 03  &   \nodata      &   \nodata  & \nodata &  84 &  \nodata \\
Nov 09  &   \nodata      &   \nodata  & \nodata &  71 &  1.42 \\ 
Avg     & \nodata         & -121 $\pm$ 8.1 & 76 $\pm$ 4.9  & 100 $\pm$ 5.2 & \nodata \\ 
\enddata
\end{deluxetable}

\begin{deluxetable}{lcccccccc}
\def\dts{\dots}
\tablewidth{0 pt}
\tabletypesize{\footnotesize}
\tablenum{A1} 
\tablecaption{SMARTS 1.3-m ANDICAM Photometry of NGC 300 OT2008-1}
\tablehead{
\colhead{HJD$-$} &
\colhead{Elapsed} &
\colhead{$B$} &
\colhead{$V$} &
\colhead{$R$} &
\colhead{$I$} &
\colhead{$J$} &
\colhead{$H$} &
\colhead{$K$} \\
\colhead{2400000} &
\colhead{days} &
\colhead{} &
\colhead{} &
\colhead{} &
\colhead{} &
\colhead{} &
\colhead{} &
\colhead{} 
}
\startdata							    
54600.646 &   0.000 & $\dts$ & $\dts$ & 14.181 & $\dts$ & $\dts$ & $\dts$ & $\dts$ \\
54601.642 &   0.996 & $\dts$ & $\dts$ & 14.231 & $\dts$ & $\dts$ & $\dts$ & $\dts$ \\
54602.936 &   2.290 & 15.495 & 14.687 & 14.211 & 13.714 & 13.052 & 12.501 & 11.435 \\
54606.910 &   6.264 & 15.525 & 14.707 & 14.251 & 13.804 & 13.048 & 12.675 & 11.694 \\
54607.939 &   7.293 & 15.535 & 14.707 & 14.261 & 13.834 & 13.065 & 12.585 & 11.643 \\
54608.904 &   8.258 & 15.595 & 14.747 & 14.271 & 13.814 & 13.042 & 12.544 & 11.717 \\
54609.902 &   9.256 & 15.655 & 14.767 & 14.291 & 13.814 & 13.109 & 12.622 & 11.675 \\
54610.873 &  10.227 & 15.615 & 14.787 & 14.271 & 13.824 & 13.093 & 12.657 & 11.470 \\
54611.886 &  11.240 & 15.655 & 14.777 & 14.291 & 13.854 & 13.062 & 12.563 & 11.781 \\
54615.934 &  15.288 & 15.745 & 14.867 & 14.331 & 13.854 & 13.043 & 12.565 & 11.708 \\
54623.911 &  23.265 & 15.915 & 14.977 & 14.381 & 13.874 & 13.027 & 12.641 & 11.607 \\
54624.879 &  24.233 & 15.965 & 14.967 & 14.391 & 13.894 & 13.102 & 12.579 & 11.659 \\
54626.877 &  26.231 & 15.985 & 15.007 & 14.411 & 13.914 & 13.106 & 12.640 & 11.676 \\
54627.876 &  27.230 & 15.995 & 15.007 & 14.411 & 13.884 & 13.093 & 12.578 & 11.759 \\
54628.899 &  28.253 & 16.015 & 15.027 & 14.411 & 13.894 & 13.017 & 12.567 & 11.677 \\
54629.835 &  29.189 & 16.055 & 15.027 & 14.441 & 13.914 & 13.117 & 12.551 & 11.911 \\
54630.878 &  30.232 & 16.075 & 15.067 & 14.431 & 13.914 & 13.081 & 12.639 & 11.887 \\
54631.861 &  31.215 & 16.095 & 15.067 & 14.431 & 13.924 & 13.143 & 12.673 & 11.741 \\
54633.888 &  33.242 & 16.115 & 15.097 & 14.451 & 13.914 & 13.043 & 12.658 & 11.477 \\
54634.863 &  34.217 & 16.195 & 15.127 & 14.471 & 13.934 & 13.069 & 12.660 & 11.625 \\
54637.847 &  37.201 & 16.215 & 15.167 & 14.511 & 13.974 & 13.167 & 12.729 & 11.592 \\
54638.907 &  38.261 & 16.295 & 15.207 & 14.541 & 13.994 & 13.140 & 12.646 & 11.896 \\
54639.891 &  39.245 & 16.335 & 15.207 & 14.591 & 14.034 & 13.315 & 12.653 & $\dts$ \\
54640.842 &  40.196 & 16.335 & 15.277 & 14.551 & 14.014 & 13.189 & 12.627 & 11.847 \\
54641.878 &  41.232 & 16.375 & 15.267 & 14.611 & 14.064 & 13.202 & 12.688 & 11.664 \\
54642.878 &  42.232 & 16.395 & 15.337 & 14.621 & 14.094 & $\dts$ & $\dts$ & $\dts$ \\
54643.868 &  43.221 & 16.415 & 15.347 & 14.671 & 14.084 & 13.193 & 12.724 & 11.964 \\
54644.876 &  44.230 & 16.455 & 15.357 & 14.661 & 14.134 & 13.167 & 12.797 & 11.960 \\
54645.882 &  45.236 & 16.495 & 15.387 & 14.691 & 14.134 & 13.188 & 12.787 & 11.999 \\
54646.799 &  46.153 & 16.505 & 15.427 & $\dts$ & $\dts$ & $\dts$ & $\dts$ & $\dts$ \\
54647.894 &  47.248 & 16.575 & 15.447 & 14.751 & 14.174 & 13.231 & 12.802 & $\dts$ \\
54648.844 &  48.198 & 16.595 & 15.487 & 14.781 & 14.214 & 13.398 & 12.964 & 12.187 \\
54649.903 &  49.257 & 16.625 & 15.517 & 14.801 & 14.234 & 13.358 & 12.888 & 12.089 \\
54650.839 &  50.193 & 16.675 & 15.547 & 14.841 & 14.244 & 13.430 & 12.941 & 12.036 \\
54651.837 &  51.191 & 16.705 & 15.597 & 14.861 & 14.284 & 13.350 & 12.914 & 11.981 \\
54652.859 &  52.213 & 16.745 & 15.627 & 14.891 & 14.314 & 13.389 & 12.929 & 12.032 \\
54653.849 &  53.203 & 16.725 & 15.657 & 14.911 & 14.334 & $\dts$ & $\dts$ & 12.225 \\
54654.887 &  54.240 & 16.845 & 15.697 & 14.951 & 14.374 & 13.504 & 12.907 & $\dts$ \\
54655.892 &  55.246 & 16.875 & 15.747 & 14.991 & 14.404 & 13.441 & 12.944 & 12.237 \\
54657.851 &  57.205 & 16.955 & 15.777 & 15.031 & 14.474 & 13.474 & 12.976 & 11.971 \\
54658.887 &  58.241 & 16.935 & 15.867 & 15.071 & 14.474 & $\dts$ & $\dts$ & $\dts$ \\
54660.879 &  60.233 & 17.095 & 15.977 & 15.181 & 14.534 & 13.545 & 13.141 & 12.151 \\
54661.883 &  61.237 & 17.115 & 15.987 & 15.201 & 14.594 & $\dts$ & $\dts$ & $\dts$ \\
54662.898 &  62.252 & 17.185 & 16.037 & 15.231 & 14.614 & 13.598 & 13.060 & 12.290 \\
54663.875 &  63.229 & 17.235 & 16.067 & 15.281 & 14.654 & 13.556 & 13.019 & 12.253 \\
54664.891 &  64.245 & 17.295 & 16.147 & 15.311 & 14.674 & 13.653 & 13.081 & 12.295 \\
54665.885 &  65.239 & 17.335 & 16.167 & 15.331 & 14.714 & 13.614 & 13.013 & 12.185 \\
54666.804 &  66.158 & 17.355 & 16.207 & 15.381 & 14.764 & 13.698 & 13.175 & 11.988 \\
54671.794 &  71.148 & 17.655 & 16.477 & 15.601 & 14.904 & 13.722 & 13.116 & 12.237 \\
54672.880 &  72.234 & 17.705 & 16.517 & 15.631 & 14.954 & 13.788 & 13.055 & 12.342 \\
54674.824 &  74.178 & 17.765 & 16.607 & 15.681 & 14.994 & 13.870 & 13.188 & 12.500 \\
54678.801 &  78.155 & 18.015 & 16.857 & 15.871 & 15.154 & 13.897 & 13.205 & 12.378 \\
54681.878 &  81.232 & 18.245 & 17.037 & 16.011 & 15.254 & 13.970 & 13.249 & 12.417 \\
54682.809 &  82.163 & 18.255 & 17.067 & 16.051 & 15.314 & 13.974 & 13.214 & 12.281 \\
54683.813 &  83.167 & 18.325 & 17.137 & 16.091 & 15.334 & 14.027 & 13.318 & 12.230 \\
54684.773 &  84.127 & 18.385 & 17.207 & 16.151 & 15.384 & 14.002 & 13.305 & 12.232 \\
54685.778 &  85.132 & 18.455 & 17.277 & 16.181 & 15.394 & 14.082 & 13.297 & 12.305 \\
54686.804 &  86.158 & 18.535 & 17.317 & 16.241 & 15.444 & 14.060 & 13.344 & 12.528 \\
54687.779 &  87.133 & 18.585 & 17.377 & 16.271 & 15.494 & 14.084 & 13.238 & 12.592 \\
54688.865 &  88.219 & 18.625 & 17.447 & 16.341 & 15.534 & 14.112 & 13.361 & 12.624 \\
54690.829 &  90.183 & 18.775 & 17.587 & 16.431 & 15.634 & 14.070 & 13.297 & 12.440 \\
54698.787 &  98.141 & 19.285 & 18.167 & 16.871 & 16.014 & 14.340 & 13.495 & 12.628 \\
54701.761 & 101.115 & 19.515 & 18.367 & 17.041 & 16.184 & 14.457 & 13.475 & 12.796 \\
54704.786 & 104.140 & 19.695 & 18.637 & 17.211 & 16.354 & 14.506 & 13.594 & 12.661 \\
54708.767 & 108.121 & 19.955 & 18.987 & 17.461 & 16.634 & 14.712 & 13.726 & 12.649 \\
54715.779 & 115.133 & 20.345 & 19.447 & 17.811 & 17.054 & 14.886 & 13.915 & 12.760 \\
54717.746 & 117.100 & 20.425 & 19.497 & 17.901 & 17.154 & 15.024 & 14.031 & 12.959 \\
54721.762 & 121.116 & $\dts$ & $\dts$ & 18.001 & 17.294 & 14.994 & 14.092 & 12.971 \\
54724.805 & 124.159 & 20.855 & 19.867 & 18.051 & 17.354 & 15.365 & 14.246 & 13.014 \\
54728.708 & 128.062 & $\dts$ & 19.837 & 18.101 & 17.494 & 15.333 & 14.224 & 13.102 \\
54732.732 & 132.086 & 21.015 & 19.927 & 18.211 & 17.634 & 15.354 & 14.288 & 13.115 \\
54735.744 & 135.098 & $\dts$ & $\dts$ & 18.231 & 17.674 & 15.582 & 14.429 & 13.299 \\
54738.711 & 138.065 & 21.075 & 20.157 & 18.331 & 17.754 & 15.784 & 14.596 & 13.187 \\
54739.748 & 139.102 & 21.075 & 20.147 & 18.311 & 17.774 & 15.742 & 14.572 & 13.302 \\
54744.712 & 144.066 & 21.185 & 20.257 & 18.421 & 17.844 & 15.777 & 14.569 & 13.304 \\
54749.713 & 149.067 & 21.195 & 20.187 & 18.451 & 17.954 & 16.032 & 14.765 & 13.255 \\
54754.711 & 154.065 & 21.155 & 20.327 & 18.531 & 18.054 & 16.131 & 14.812 & 13.361 \\
54760.708 & 160.062 & 21.345 & 20.487 & 18.571 & 18.114 & 16.282 & 14.832 & 13.413 \\
54766.725 & 166.079 & 21.565 & 20.547 & 18.681 & 18.214 & 16.356 & 14.999 & 13.457 \\
54772.674 & 172.028 & 21.525 & 20.727 & 18.711 & 18.304 & 16.297 & 15.104 & 13.561 \\
54778.644 & 177.998 & 21.625 & 20.787 & 18.791 & 18.394 & 16.496 & 15.067 & 13.570 \\
54784.671 & 184.025 & 21.725 & 20.967 & 18.821 & 18.524 & 16.612 & 15.181 & 13.565 \\
54794.651 & 194.005 & 21.795 & 20.957 & 18.941 & 18.574 & 16.681 & 15.322 & 13.593 \\
54801.645 & 200.999 & 22.095 & 21.127 & 18.941 & 18.614 & 16.861 & 15.370 & 13.709 \\
54807.643 & 206.997 & $\dts$ & $\dts$ & 18.991 & 18.674 & $\dts$ & 15.312 & 13.808 \\
54817.602 & 216.956 & $\dts$ & 21.147 & 19.021 & 18.814 & $\dts$ & 15.499 & 13.754 \\
54827.608 & 226.962 & $\dts$ & 21.367 & 19.171 & 18.874 & $\dts$ & $\dts$ & $\dts$ \\  
54839.608 & 238.962 & $\dts$ & $\dts$ & 19.231 & 18.994 & $\dts$ & $\dts$ & $\dts$ \\	
54845.543 & 244.897 & $\dts$ & $\dts$ & 19.371 & $\dts$ & $\dts$ & $\dts$ & $\dts$ \\	
54847.563 & 246.917 & $\dts$ & 21.687 & $\dts$ & $\dts$ & $\dts$ & $\dts$ & $\dts$ \\	
54848.546 & 247.900 & $\dts$ & $\dts$ & 19.331 & $\dts$ & $\dts$ & $\dts$ & $\dts$ \\	
54849.535 & 248.889 & $\dts$ & $\dts$ & $\dts$ & 19.174 & $\dts$ & $\dts$ & $\dts$ \\	
54851.557 & 250.911 & $\dts$ & 21.717 & $\dts$ & $\dts$ & $\dts$ & $\dts$ & $\dts$ \\
54852.540 & 251.894 & $\dts$ & $\dts$ & 19.371 & $\dts$ & $\dts$ & $\dts$ & $\dts$ \\
54853.534 & 252.888 & $\dts$ & $\dts$ & $\dts$ & 19.144 & $\dts$ & $\dts$ & $\dts$ \\
54855.554 & 254.908 & $\dts$ & 21.767 & $\dts$ & $\dts$ & $\dts$ & $\dts$ & $\dts$ \\
54857.530 & 256.884 & $\dts$ & $\dts$ & $\dts$ & 19.234 & $\dts$ & $\dts$ & $\dts$ \\
54861.529 & 260.883 & $\dts$ & $\dts$ & $\dts$ & 19.294 & $\dts$ & $\dts$ & $\dts$ \\
54872.532 & 271.886 & $\dts$ & $\dts$ & 19.521 & $\dts$ & $\dts$ & $\dts$ & $\dts$ \\
54873.529 & 272.883 & $\dts$ & $\dts$ & $\dts$ & 19.324 & $\dts$ & $\dts$ & $\dts$ \\
54962.912 & 362.266 & $\dts$ & $\dts$ & $\dts$ & $\dts$ & 19.675 & $\dts$ & $\dts$ \\
54964.920 & 364.274 & $\dts$ & $\dts$ & 21.281 & $\dts$ & $\dts$ & 16.233 & $\dts$ \\
54965.887 & 365.241 & $\dts$ & $\dts$ & $\dts$ & $\dts$ & $\dts$ & $\dts$ & 14.389 \\
54975.905 & 375.259 & $\dts$ & $\dts$ & $\dts$ & $\dts$ & 19.240 & 16.386 & 14.157 \\
54982.905 & 382.259 & $\dts$ & $\dts$ & $\dts$ & $\dts$ & 20.536 & 16.642 & 14.445 \\
54995.883 & 395.237 & $\dts$ & $\dts$ & $\dts$ & $\dts$ & 20.545 & 16.998 & 14.485 \\
55005.908 & 405.262 & $\dts$ & $\dts$ & 22.191 & 22.544 & 20.834 & 16.820 & 14.473 \\
55014.914 & 414.268 & $\dts$ & $\dts$ & $\dts$ & $\dts$ & 20.493 & 17.194 & 14.394 \\
55036.822 & 436.176 & $\dts$ & $\dts$ & $\dts$ & $\dts$ & 20.760 & $\dts$ & $\dts$ \\
\enddata
\tablecomments{First two lines are Bronberg discovery observations, taken to be
$R$ magnitudes. All other data from 1.3-m ANDICAM frames. Typical errors are
$\pm$0.02--0.04~mag until end of 2008 (HJD$\simeq$54830), increasing thereafter
to approximately $\pm$0.1--0.2~mag as the transient faded. }
\end{deluxetable}

\begin{deluxetable}{lrrl}
\tabletypesize{\footnotesize}
\tablenum{A2}
\tablecaption{Journal of  Spectroscopic Observations}
\tablewidth{0pt}
\tablehead{
\colhead{Date (UT)} &  \colhead{Spec/Grating} & \colhead{Integration (sec)} & \colhead{Comment} 
  }
  \startdata
  2008 May 15 &  RC 13  &     3x 300  &       \\   
  2008 May 16 &  RC 26  &     3x 600  &      no signal\\  
  2008 May 17 &  RC 9   &    3x 600   &                \\  
  2008 Jun 20 &  RC 13  &      3x 600 &               \\   
  2008 Jun 21 &  RC  47 &       3x 600 &              \\   
  2008 Jun 22 &  RC  26 &      3x 900 &       only H-beta detected  \\  
  2008 Jul 06 &  RC 58  &     3x 800  &                \\   
  2008 Jul 06 &  MagE/echellette &  2x 800   &            \\
  2008 Jul 08 &  RC 58  &     3x 600  &               \\   
  2008 Jul 18 &  RC 13  &     3x 600  &                \\  
  2008 Aug 01 &  RC 58  &      3x 600  &              \\   
  2008 Aug 06 &  RC 26  &      3x1200  &               \\   
  2008 Aug 07 &  RC 58  &      3x 600  &               \\   
  2008 Aug 23 &  RC 13  &      3x 900  &                \\   
  2008 Aug 24 &  RC 58  &      3x1000  &                \\    
  2008 Aug 25 &  RC 47  &      3x1100  &                \\   
  2008 Aug 30 & MagE/echellette  &  900 &            \\
  2008 Sep 01 & MagE/echellette  &  900 &            \\ 
  2008 Sep 08 &  RC 13  &      3x1000  &                \\    
  2008 Sep 10 &  RC  47 &      3x1200  &                 \\   
  2008 Sep 13 &  RC 47   &      3x1200  &                 \\
  2008 Sep 16 &  RC 47   &      3x1200  &                 \\
  2008 Sep 21 & RC 58   &      3x 600  &                \\
  2008 Sep 27 & RC 47   &      3x1200  &                \\
  2008 Sep 28 & RC 26   &      3x1200  &                \\
  2008 Oct 01 & RC 13   &      3x 900  &                \\
  2008 Oct 03 & RC 47   &      3x1200  &                \\
  2008 Oct 12 & RC 47   &      3x1200  &                \\
  2008 Oct 13 & RC 13   &      3x1200  &                 \\
  2008 Nov 03 & RC 47   &      3x1200  &                \\
  2008 Nov 04 & RC 13   &      3x1200  &                \\
  2008 Nov 07 & RC 58   &      3x 900  &                \\
  2008 Nov 09 & RC 47   &      3x1200  &                \\
  2008 Nov 13 & RC 13   &      3x1200  &                 \\
  2008 Dec 01 & RC 47   &      3x1200  &                \\
  2008 Dec 06 & RC 13   &      3x1100  &                \\
  2008 Dec 19 & RC 58   &      3x 900  &               \\ 
  2008 Dec 30 & RC 13   &      3x1200  &                \\
  2009 Jan 16 & RC 47   &      3x1200  &                \\
  2009 Jun 05 &  MagE/echellette &  3000 &              \\ 
  2009 Jul 19 -- 23 & ISAAC &          &  1.65$\mu$m, 2.2$\mu$m, 3.55$\mu$m \\

  \enddata
  \end{deluxetable}


\begin{figure}
\figurenum{1}
\epsscale{1.0}
\plotone{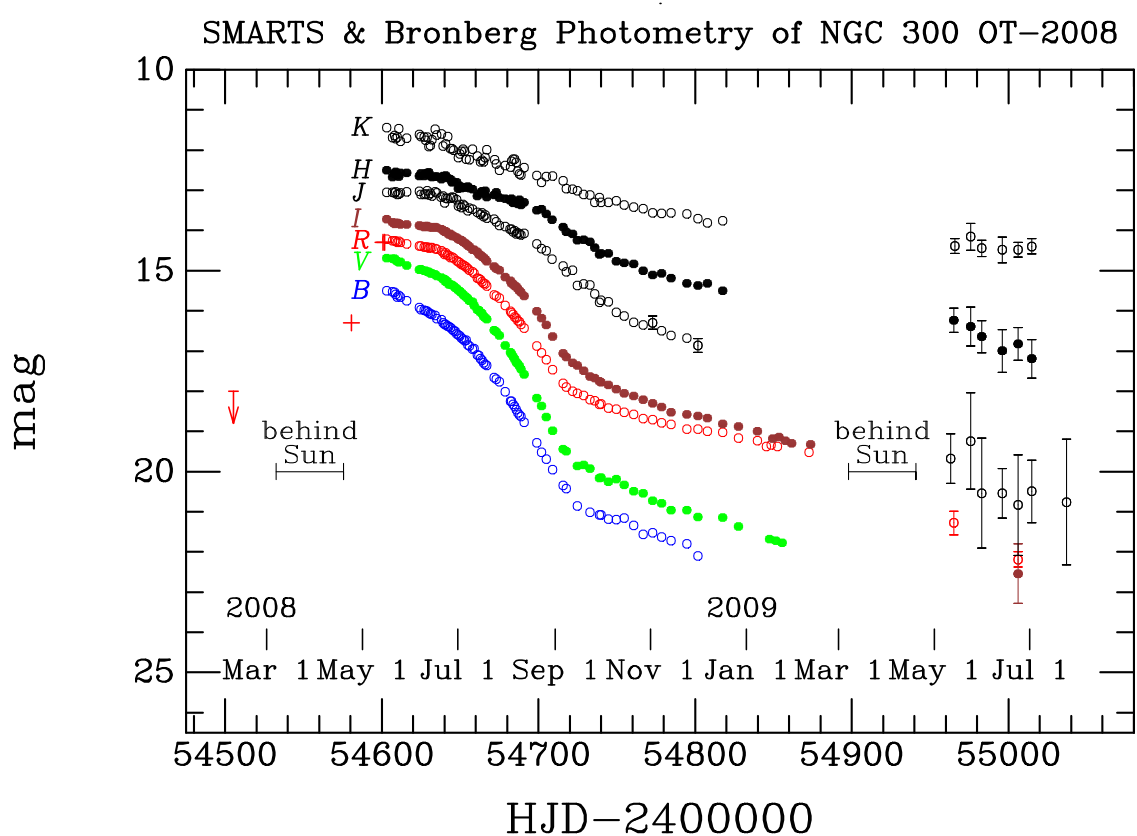}
\caption{{\it BVRIJHK\/} light curve of NGC~300 OT~2008.
SMARTS 1.3-m data are shown as open and filled circles; error bars are plotted
only when larger than the plotting symbols. Bronberg discovery observations and
the pre-discovery detection are shown as crosses (these broad-band magnitudes
are close to Landolt $R$), and the downward arrow on the left shows the Bronberg
upper limit in 2008 February. To appear in color in the electronic edition.} 
\end{figure}

\begin{figure}
\figurenum{2}
\epsscale{1.0}
\plotone{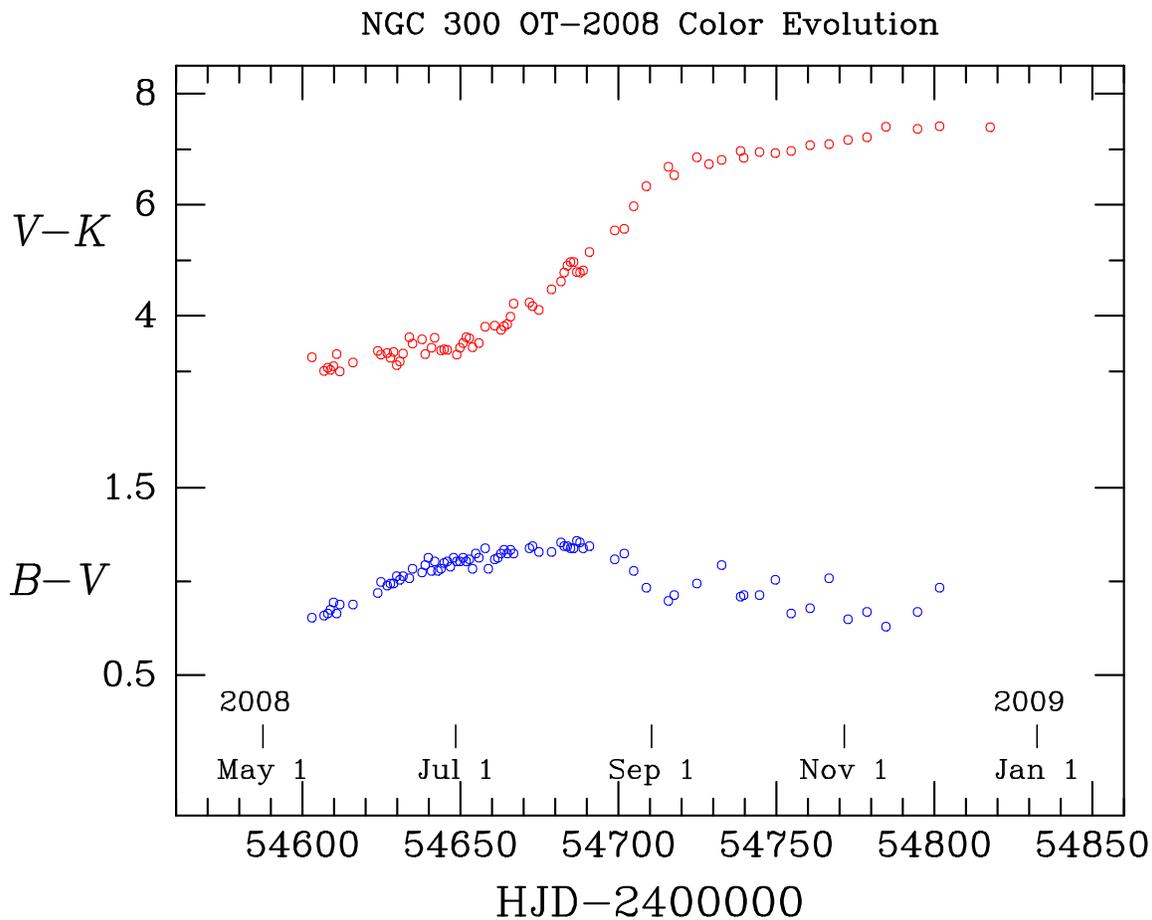}
\caption{The $B-V$ and $V-K$ colors of NGC~300 OT~2008 vs.\ time. The $V-K$
color (top points) became dramatically redder as the outburst proceeded,
evolving from 3.1 in mid-May 2008 to $\sim$7.4 at the end of our coverage. The
$V-K$ slope changed markedly in 2008 September, due to the change in slope of
$V$ vs.\ time. As discussed in the text (\S 3), $B-V$ also became progressively
redder with time, until 2008 August. It then became {\it bluer\/} with time
until the end of our coverage. To appear in color in the electronic edition.}
\end{figure}

\begin{figure}
\figurenum{3}
\epsscale{.80}
\plotone {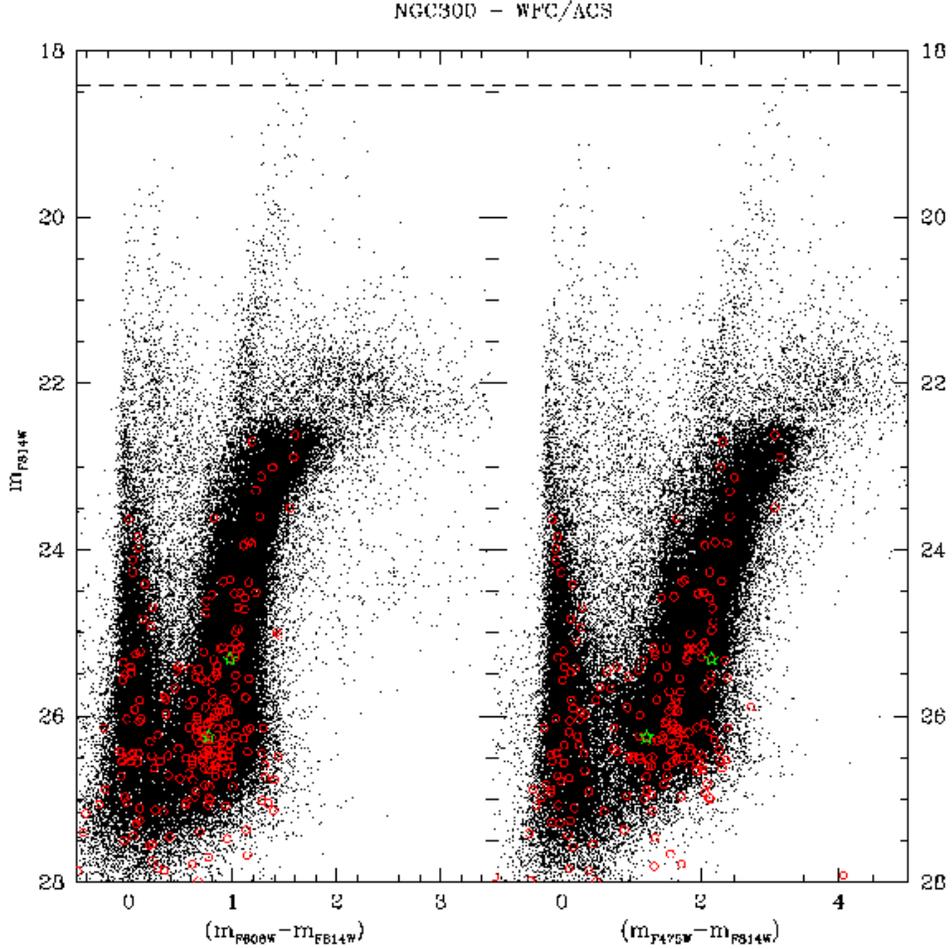}
\caption{Deep CMDs derived from the archical {\it HST}/ACS and WFC images, as
described in the text. The left frame shows $I$ vs.\ $V-I$, and the right frame
shows $I$ vs.\ $B-I$\null. Black dots show all point sources with reliable
photometry. The red circles mark stars within a radius of $2\farcs5$ (23~pc)
from the NGC~300~OT.  The two green stars are the two only stars lying within
$0\farcs25$ (2.3~pc) from the site of the OT, but well outside the astrometric
error box.}
\label{cmds}
\end{figure}

\begin{figure}
\figurenum{4}
\epsscale{1.2}
\plottwo{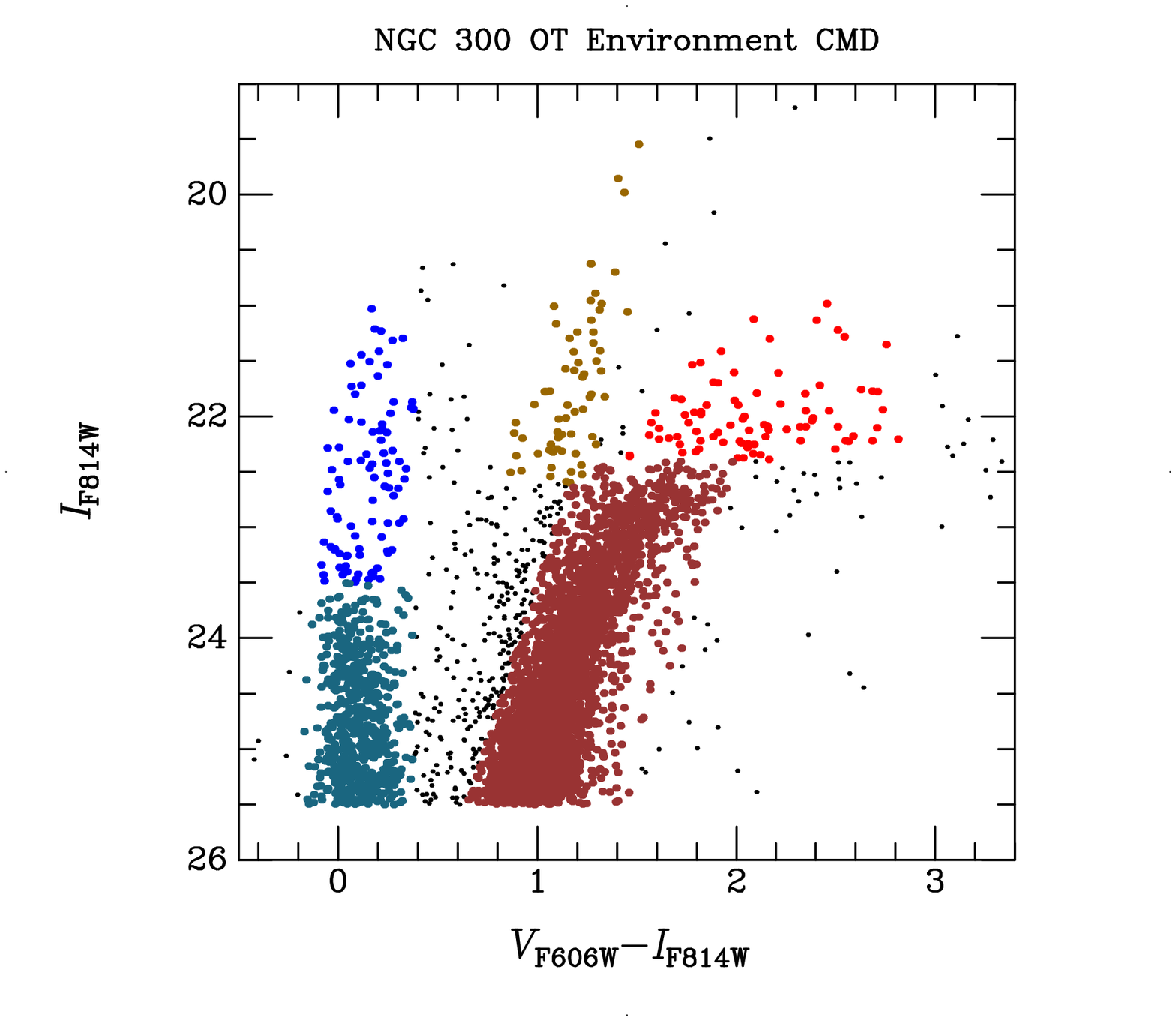}{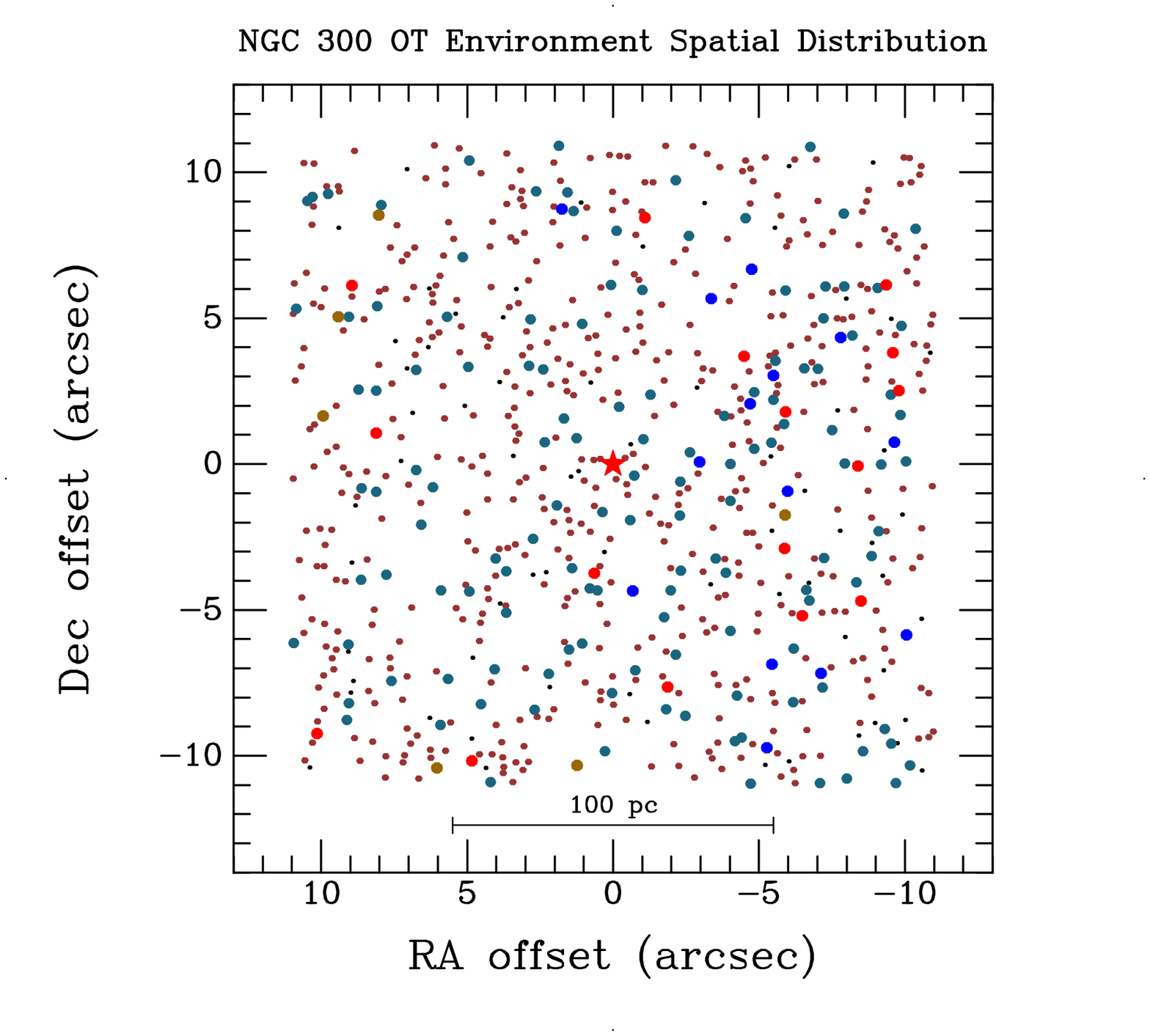}
\caption{Left: The color-magnitude diagram for stars lying within a
$500\times500$~pc square centered on the NGC~300 2008 OT. Color-coding;  bright
blue: bright main-sequence and blue-loop stars; slate blue: faint main-sequence
stars; orange: red supergiants; bright red: AGB stars; dark red: red giants;
black: all other stars in the field. Right: The spatial distribution of stars in
a $200\times200$~pc square centered on the OT (represented by the large red
star); same color-coding as in the left panel is used.}
\end{figure}

\begin{figure}
\figurenum{5}
\epsscale{1.0}
\plotone{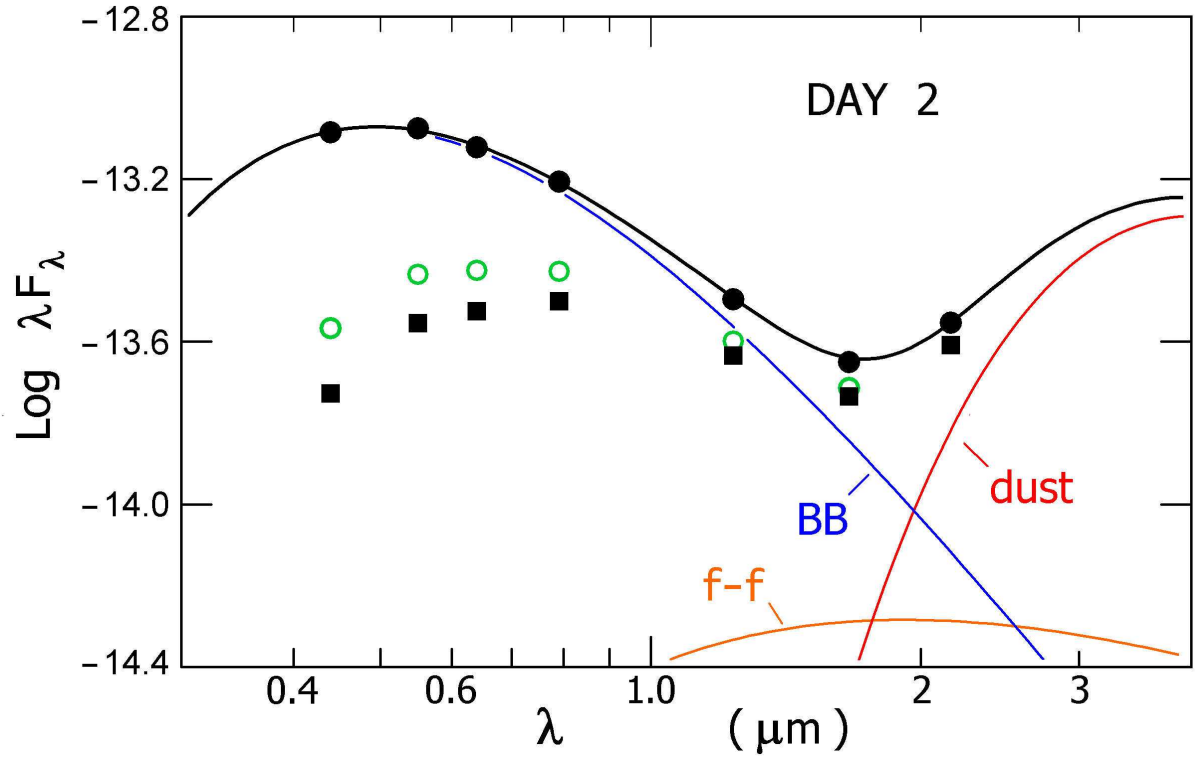}
\caption{The spectral energy distribution, log Watts m$^{2}$ vs.\ log $\lambda$, 
at maximum light. The observed photometry is plotted as filled squares. The
magnitudes corrected for the mean $A_{V}$ and for $A_{V} = 1.2$ mag are shown as
open and filled circles, respectively. The 7500 K blackbody for the latter
points, the free-free contribution, and the 715 K dust are plotted separately.
The  combined contribution, the SED,  is shown as a solid black line.}
\end{figure} 

\begin{figure}
\figurenum{6}
\epsscale{1.1}
\plottwo{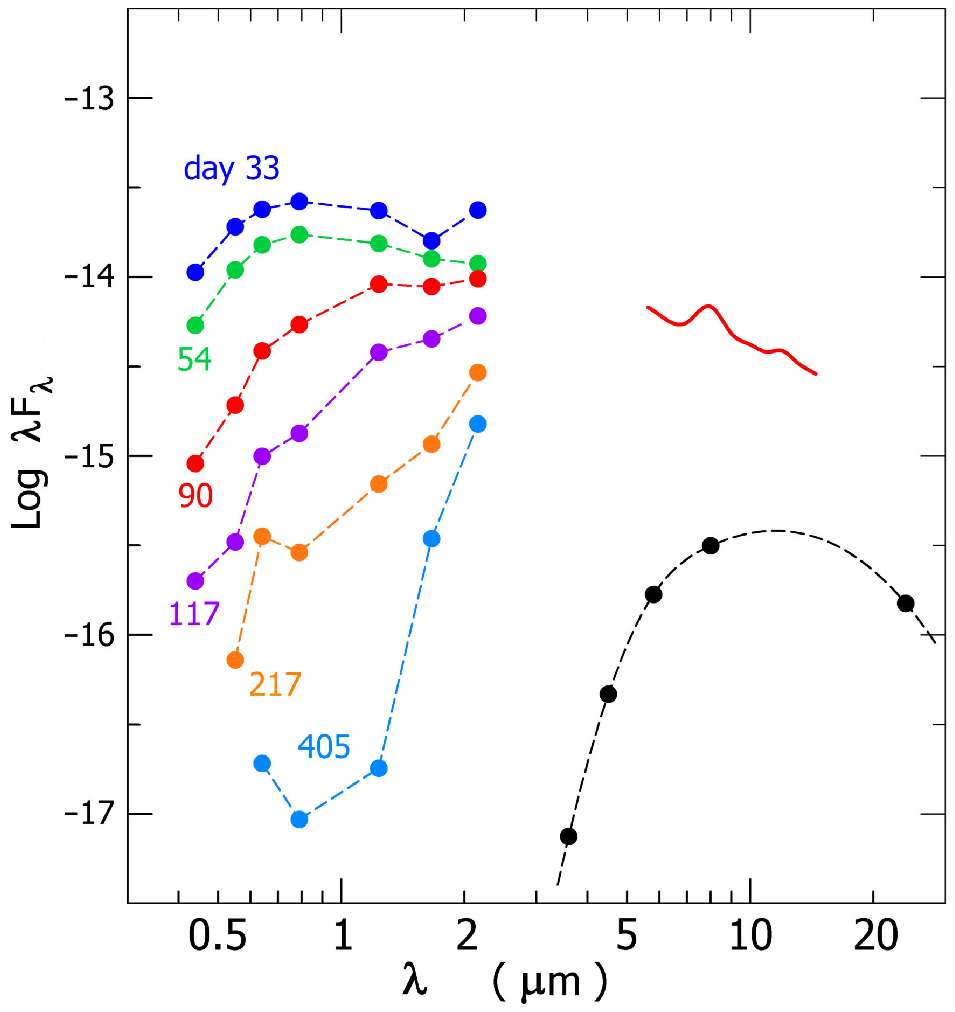}{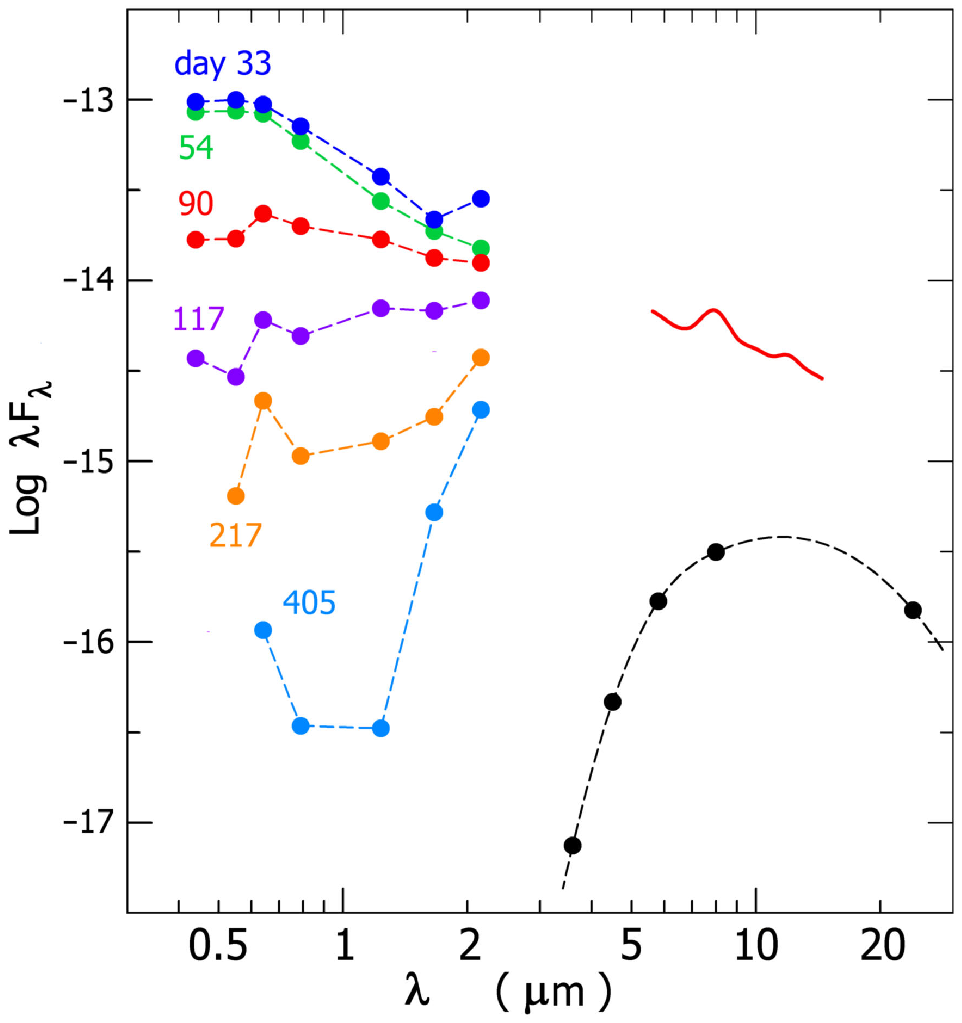}
\caption{The evolution of the spectral energy distribution (log Watts m$^{2}$
vs. log $\lambda$) from day 33 to our last photometric observation with $R$ and
$I$ magnitudes on 2009 June 6 (day 405). The observed magnitudes are plotted in 
the left panel and the SEDs corrected for extinction, as described in the text,
are  shown in the right  panel. The mid-IR emission  feature is shown with the
day 90 photometry.  The pre-eruption SED, in black,  is also shown for
comparison. The peak at $R$ is due to strong H$\alpha$ emission. The extinction
corrections used for the SEDs in the right panel are: day 33, $A_{V} = 1.8$, day
54, $A_{V} = 2.25$, day 90, $A_{V} = 2.37$. }
\end{figure}

\begin{figure}
\figurenum{7}
\epsscale{1.0}
\plotone{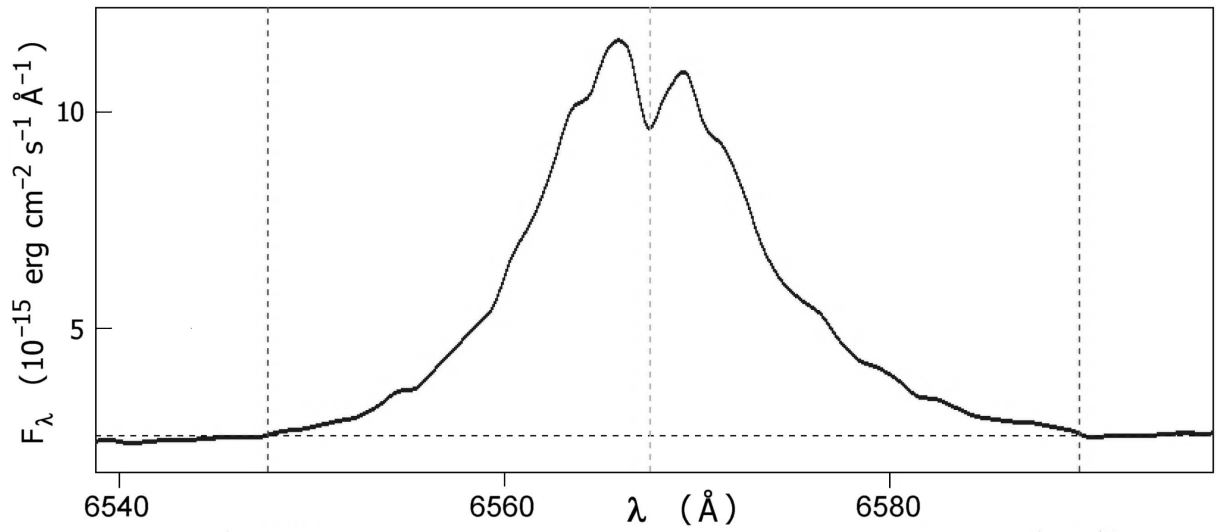}
\caption{The H$\alpha$ line from 2008 July 6  showing the Thomson scattering 
wings. The dashed lines mark the blue and red extent of the wings and the center
of the absorption minimum, see text. Also note the split profile with multiple
components discussed in the text.}
\end{figure}

\begin{figure}
\figurenum{8}
\epsscale{1.0}
\plotone{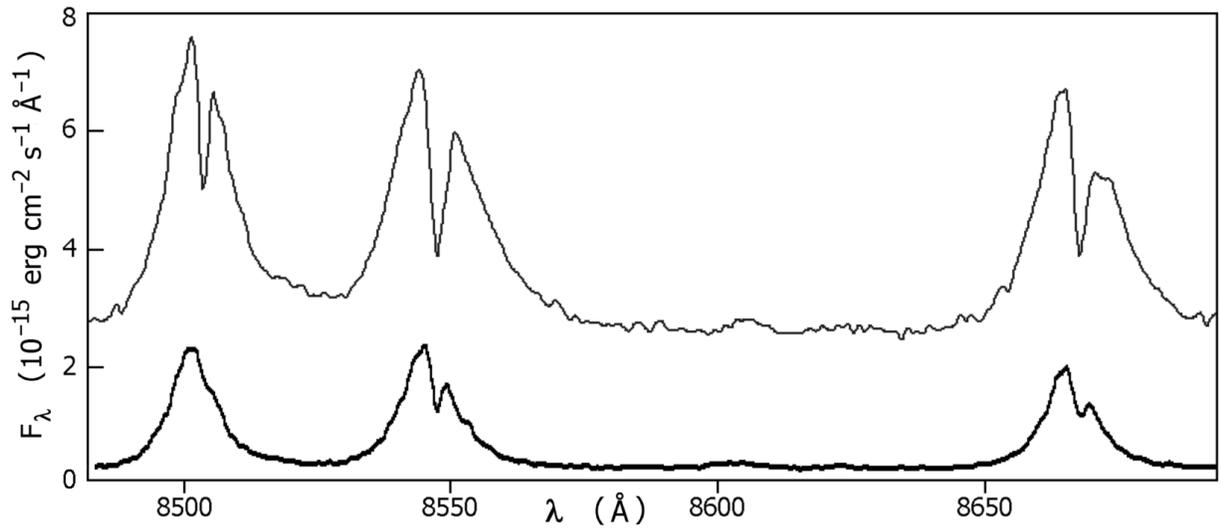}
\caption{The \ion{Ca}{2} triplet from 2008 July 6 (top)  and  August
30--September 1 (bottom), showing the evidence for multiple components and the
variation in the profiles. The red emission component in the $\lambda$8498~{\AA}
line is no longer present in the echellette  spectra from August 30  and
September 1 and there is no clear absorption minimum, although a shoulder to the
blue and red of the emission peak indicates the presence of additional
emission.  The other two lines in the triplet still  show well-resolved double
profiles. Also note the prominent redward wings similar to the [\ion{Ca}{2}]
lines in Figure 10.}
\end{figure}

\begin{figure}
\figurenum{9}
\epsscale{1.0}
\plotone{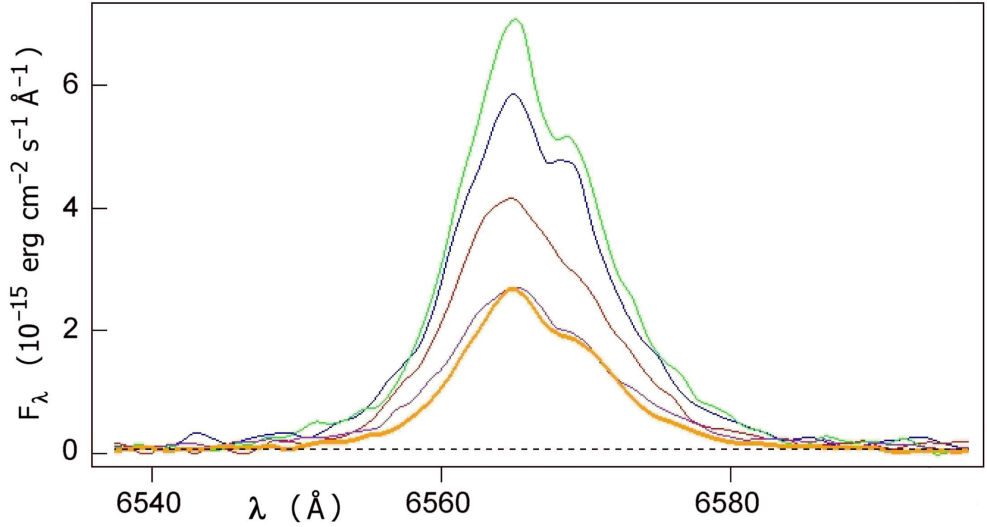}
\caption{Variation in the appearance of the H$\alpha$ double-peaked profiles
from 2008 Sep 10  to 2008  Oct 13. The profiles are color-coded by date: blue:
Sep 10, red: Sep 13, green: Sep 16, purple: Sep 27, and gold: Oct 13.} 
\end{figure}

\begin{figure}
\figurenum{10}
\epsscale{1.0}
\plotone{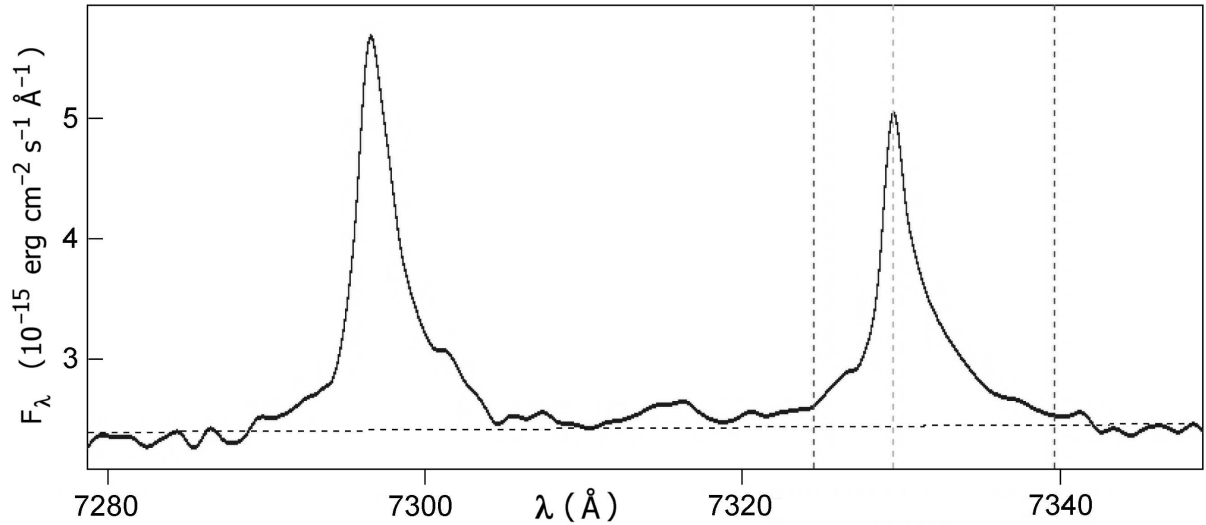}
\caption{The [\ion{Ca}{2}] lines from 2008 July 6, showing the asymmetric
profiles and the Thomson scattering wings. The dashed lines mark the blue and
red extent of the wings on the 7323~{\AA} line.}
\end{figure}

\begin{figure}
\figurenum{11}
\epsscale{1.0}
\plotone{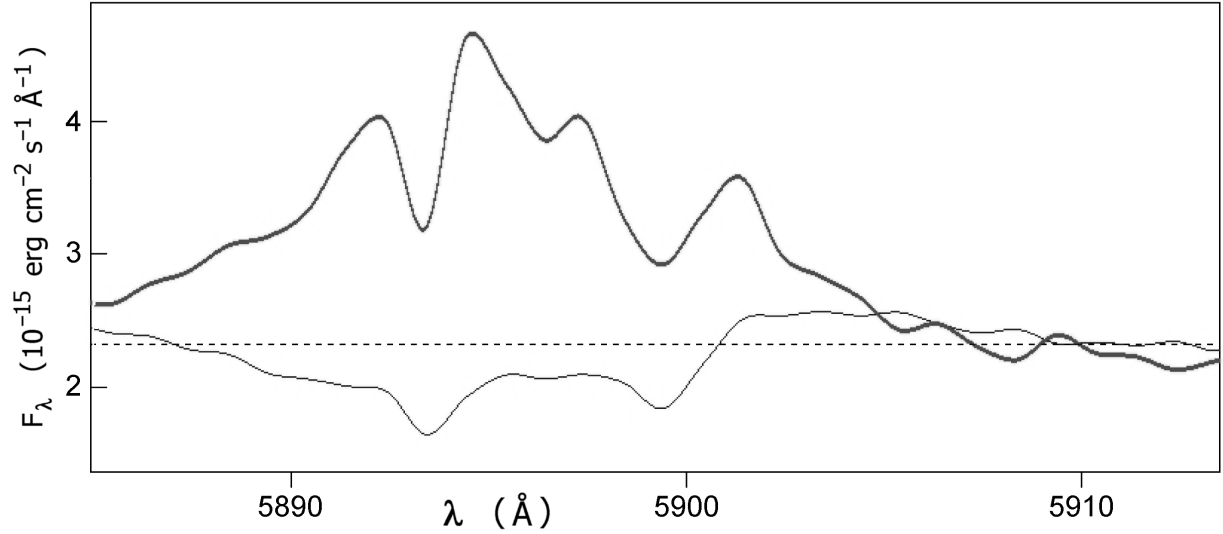}
\caption{The \ion{Na}{1} D lines transition from a pure absorption profile on
2008 Jul 6, the lower profile, to an emission profile with absorption on 2008
Aug 30, the upper, darker profile.}
\end{figure}

\begin{figure}
\figurenum{12}
\epsscale{1.0}
\plotone{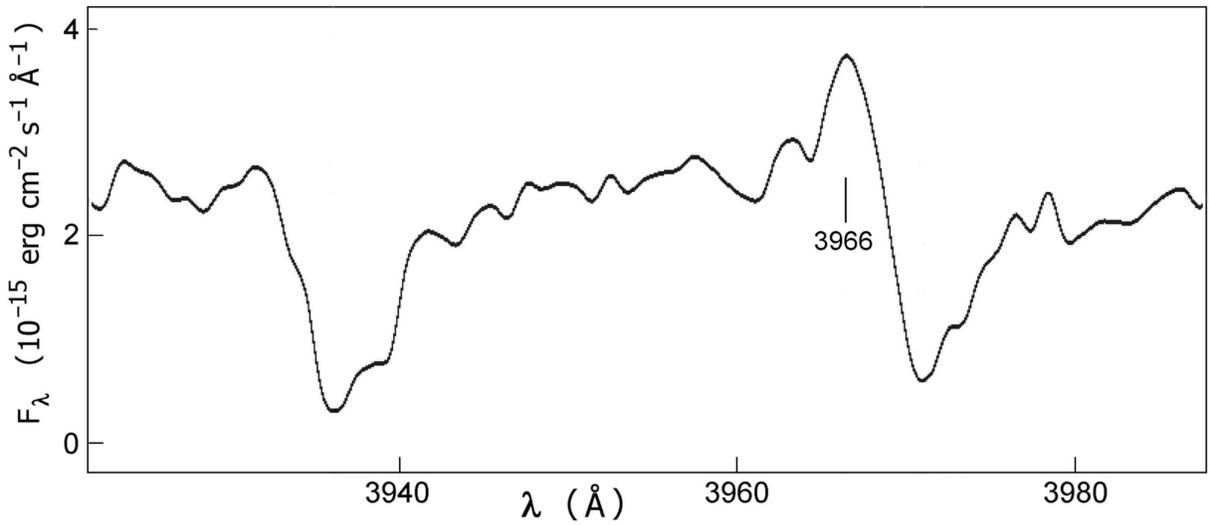}
\caption{The \ion{Ca}{2} H and K absorption lines from 2008 Jul 6 showing the
double  absorption profiles in each line. The emission line at 3966{\AA} is a
blend of He I $\lambda$3964 and Fe II $\lambda$3964.}
\end{figure} 

\begin{figure}
\figurenum{13}
\epsscale{1.0}
\plotone{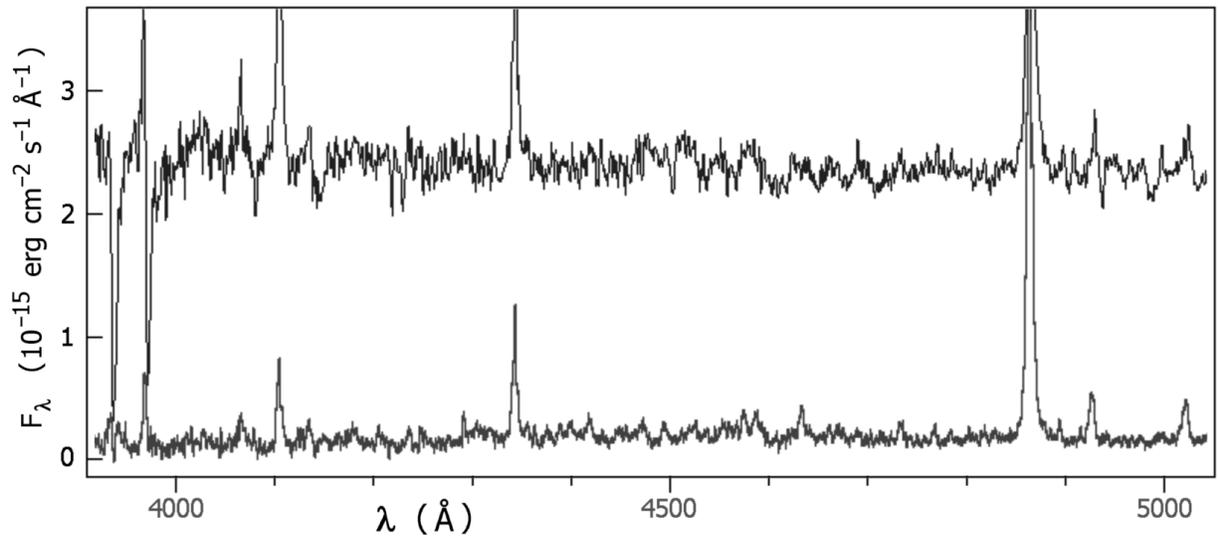}  
\caption{The echelette spectra in the blue from approximately \ion{Ca}{2} H and
K to 5000{\AA} illustrating the change in the absorption line spectrum from 2008
Jul 6, upper  spectrum,  to 2008  Aug 30/Sep 01, the lower spectrum.} 
\end{figure}  

\begin{figure}
\figurenum{14}
\epsscale{1.0}
\plotone{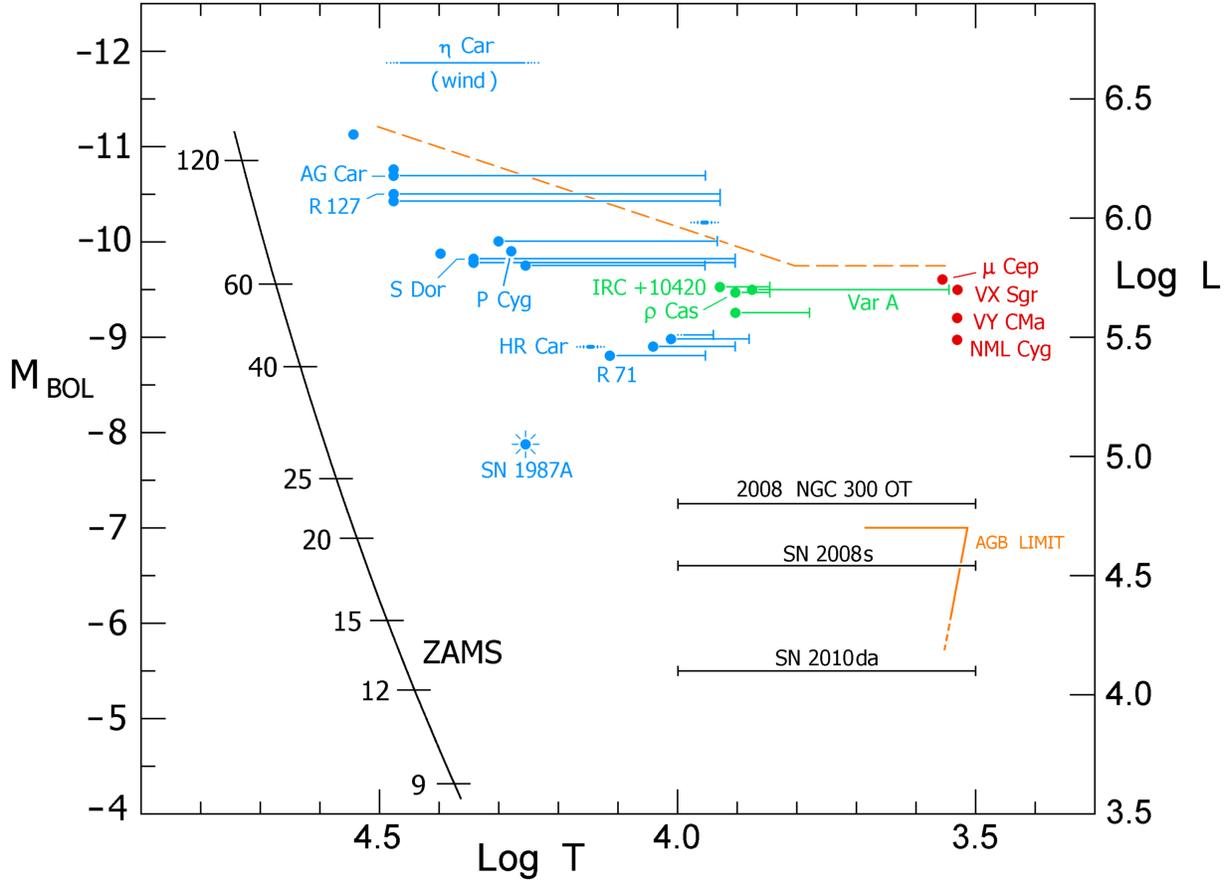}
\caption{A schematic HR diagram showing the locations of the progenitors of 
NGC 300 OT2008-1, SN~2008S and SN~2010da. Several well-known unstable and
high-mass-losing stars are also shown, including $\eta$~Car and known LBVs, in
blue with their  transits in apparent temperatures during their optically thick
wind stage or eruption, and the warm and cool hypergiants, in green and red,
respectively.  The empirical upper luminosity boundary is shown as a dashed line
(see Humphreys \& Davidson 1994).}
\end{figure}


\end{document}